\documentclass[a4paper,11pt]{article}
\usepackage{amssymb}

\usepackage{graphicx}
\usepackage{amsmath}

 \DeclareFontFamily{U}{rsf}{}
\DeclareFontShape{U}{rsf}{m}{n}{
  <5> <6> rsfs5 <7> <8> <9> rsfs7 <10-> rsfs10}{}
\DeclareMathAlphabet\Scr{U}{rsf}{m}{n} \makeatletter
\@addtoreset{equation}{section} \makeatother

\def\be{\begin{equation}}
\def\ee{\end{equation}}
\def\ba{\begin{array}}
\def\ea{\end{array}}

\newcommand{\bea}{\begin{eqnarray}}
\newcommand{\eea}{\end{eqnarray}}

\begin{document}

\begin{titlepage}
 \thispagestyle{empty}
\begin{flushright}
     \hfill{CERN-PH-TH/2011-246}\\
 \end{flushright}

 \vspace{70pt}

 \begin{center}
     { \Huge{\bf      {On Invariant Structures\\\vspace{10pt}of Black Hole Charges}}}

     \vspace{25pt}

     {\Large {Sergio Ferrara$^{a,b,c}$, Alessio Marrani$^{a}$ and Armen Yeranyan$^{b,d}$}}

     \vspace{40pt}

  {\it ${}^a$ Physics Department, Theory Unit, CERN,\\
     CH -1211, Geneva 23, Switzerland;\\
     \texttt{sergio.ferrara@cern.ch}\\
     \texttt{alessio.marrani@cern.ch}}

     \vspace{10pt}

    {\it ${}^b$ INFN - Laboratori Nazionali di Frascati,\\
     Via Enrico Fermi 40, I-00044 Frascati, Italy\\
     \texttt{ayeran@lnf.infn.it}}

     \vspace{10pt}

     {\it ${}^c$  Department of Physics and Astronomy,\\
University of California, Los Angeles, CA 90095-1547,USA}\\

     \vspace{10pt}

{\it ${}^d$ Department of Physics, Yerevan State University\\
Alex Manoogian St. 1, Yerevan, 0025, Armenia}\\

     \vspace{10pt}

     \vspace{60pt}

     {ABSTRACT}

 \vspace{10pt}
 \end{center}
We study ``minimal degree'' complete bases of duality- and
``horizontal''- invariant homogeneous polynomials in the flux
representation of two-centered black hole solutions in two classes
of $D=4$ Einstein supergravity models with symmetric vector
multiplets' scalar manifolds. Both classes exhibit an $SL\left(
2,\mathbb{R}\right) $ ``horizontal'' symmetry which mixes the two
centers.

The first class encompasses $%
\mathcal{N}=2$ and $\mathcal{N}=4$ matter-coupled theories, with
semi-simple $U$-duality given by $SL\left( 2,\mathbb{R}\right)
\times SO\left( m,n\right) $; the analysis is carried out in the
so-called \textit{Calabi-Vesentini} symplectic frame (exhibiting
maximal manifest covariance) and until order six in the fluxes
included.

The second class, exhibiting a non-trivial ``horizontal" stabilizer
$SO(2)$, includes $\mathcal{N}=2$ \textit{minimally coupled} and
$\mathcal{N}=3$
matter coupled theories, with $U$-duality given by the pseudo-unitary group $%
U\left( r,s\right) $ (related to complex flux representations).

Finally, we comment on the formulation of special K\"{a}hler geometry in terms of \textit{%
``generalized''} groups of type $E_{7}$.

\end{titlepage}
\tableofcontents
\section{\label{Intro}Introduction}

%\texttt{Introductory starting considerations to be filled...\medskip }

We consider \textit{reducible} \textit{symmetric} supergravity models in $%
D=4 $ space-time dimensions\footnote{%
Marginal stability for these models was studied \textit{e.g.} in \cite
{Sen,David-1}.}, which we will here dub \textit{Calabi-Vesentini} (\textit{%
CV)} models, for reasons which will be evident from treatment below. As also
given by Table 1, these models are characterized by the following $U$-duality%
\footnote{%
Here $U$-duality is referred to as the ``continuous'' symmetries of \cite
{CJ-1}. Their discrete versions are the $U$-duality non-perturbative string
theory symmetries introduced by Hull and Townsend \cite{HT-1}.} and
``horizontal'' \cite{FMOSY-1} symmetries:
\begin{equation}
\begin{array}{l}
U\text{-duality}:G_{4}=SL_{v}\left( 2,\mathbb{R}\right) \times SO\left(
m,n-2\right) ; \\
\\
\text{``horizontal''}:\mathcal{G}_{p}=SL_{h}\left( p,\mathbb{R}\right) ,
\end{array}
~~~m=\left\{
\begin{array}{l}
2~\left( \mathcal{N}=2,~n\geqslant 3\right) ; \\
\\
6~\left( \mathcal{N}=4,~n\geqslant 2\right) ,
\end{array}
\right.  \label{G4-red}
\end{equation}
where $p$ denotes the number of centers of the multi-centered solution under
consideration.

Considering an array of $p$ charge vectors $\mathcal{Q}_{a}^{M}$ ($a=1,...,p$%
) pertaining to a $p$-centered solution, the $U$-duality group acts on the
index $M$ in a symplectic representation:
\begin{equation}
\mathcal{Q}_{a}^{M}\rightarrow \mathcal{S}_{P}^{M}\mathcal{Q}_{a}^{P},~~%
\mathcal{S}^{T}\mathbb{C}\mathcal{S}=\mathbb{C},
\end{equation}
where $\mathbb{C}_{MN}$ is the symplectic-invariant metric (defined in (\ref
{C-structure}) below). On the other hand, the ``horizontal'' symmetry \label%
{FMOSY-1} acts on the index $a$ as a linear transformation on the $p$
vectors:
\begin{equation}
\mathcal{Q}_{a}^{M}\rightarrow L_{a}^{b}\mathcal{Q}_{b}^{M},~~L\in
SL_{h}\left( p,\mathbb{R}\right) .
\end{equation}
The ``horizontal'' symmetry, which is not a symmetry of the Lagrangian
formulation of the theory, proves to be useful in the classification of
multi-charge orbits, which are relevant for the dynamics of multi-centered
(black hole) solutions in supergravity \cite{FMOSY-1,ADFMT-1,Small-1}. For
the two-centered case ($p=2$) considered in the present investigation, the
lowest-order duality- and ``horizontal''- invariant polynomial is of order $%
2 $ in the charges, and it is nothing but the usual Schwinger symplectic
product $\mathcal{W}$ of two dyonic charge vectors (see (\ref{15-bis})
below).

As evident from (\ref{G4-red}), we anticipate that the case of $\mathcal{N}%
=4 $ theory coupled to $n_{V,\mathcal{N}=4}=n-2\geqslant 0$ matter (vector)
multiplets can be recovered by shifting $n\rightarrow n+4$ in all formul\ae\
of the treatment below. The semi-simple nature of $G_{4}$ justifies the name
\textit{``reducible''}, whereas \textit{``symmetric''} is due to the fact
that the corresponding scalar manifolds belong to the sequence $\mathcal{ST}%
\left[ m,n\right] $, of particular relevance for superstring
compactifications (see \textit{e.g.} the analysis in Sec. 3.1 and App. C of
\cite{N=2-Big}, and Refs. therein).

Let us now reconsider the ``$T$-tensor formalism'' for CV models, introduced
in Secs. 3 and 4 of \cite{FMOSY-1}, which will be further extended, until
order $6$ included, in Sec. \ref{p=2-CV}. A key feature of CV models is the
fact that the electro-magnetic splitting
\begin{equation}
\mathcal{Q}^{M}\equiv \left( p^{\Lambda },q_{\Lambda }\right)
\end{equation}
of the symplectic vector of the $2$-form field strengths' fluxes (also named
magnetic and electric charges) can be implemented with \textit{full manifest
covariance} with respect to $G_{4}$ (\ref{G4-red}). Namely, $\mathcal{Q}$
sits in the $\left( \mathbf{2},\mathbf{m+n-2}\right) $ bi-fundamental irrep.
of $G_{4}$, and it is thus an electro-magnetic doublet $\mathbf{2}$ of the
\textit{``vertical''} $SL_{v}\left( 2,\mathbb{R}\right) $; the symplectic
index $M$ thus splits as follows (\textit{cfr.} Eq. (3.7) of \cite{FMOSY-1})
\begin{equation}
\left.
\begin{array}{l}
M=\alpha \Lambda , \\
\alpha =1,2,~\Lambda =1,...,m+n-2.
\end{array}
\right\} \Rightarrow \mathcal{Q}^{M}\equiv \mathcal{Q}_{\alpha }^{\Lambda },
\label{CV-split}
\end{equation}
and it should be pointed out that in the $\mathcal{N}=2$ case usually $%
\Lambda =0,1,...,n-1$, with ``$0$'' pertaining to the $D=4$ graviphoton
vector. The manifestly $G_{4}$-covariant symplectic frame (\ref{CV-split})
is usually dubbed \textit{Calabi-Vesentini }frame \cite{CV}, and it was
firstly introduced in supergravity in \cite{CDFVP-1}.

\begin{table}[h]
\begin{center}
\begin{tabular}{|c||c|c|c|}
\hline
$
\begin{array}{c}
\\
\mathcal{N} \\
~
\end{array}
$ & $\frac{G_{4}}{mcs\left( G_{4}\right) }$ & $\mathit{rank}$ & $
\begin{array}{c}
\\
J_{3} \\
\mathit{reducible}
\end{array}
$ \\ \hline\hline
$
\begin{array}{c}
\\
2 \\
~
\end{array}
$ & $\frac{SL_{v}\left( 2,\mathbb{R}\right) }{U\left( 1\right) }\times \frac{%
SO\left( 2,n-2\right) }{SO\left( 2\right) \times SO\left( n-2\right) }%
,~n\geqslant 3~$ & $1+\text{min}\left( 2,n-2\right) $ & $\mathbb{R}\oplus
\mathbf{\Gamma }_{1,n-3}~$ \\ \hline
$
\begin{array}{c}
\\
4 \\
~
\end{array}
$ & $\frac{SL_{v}\left( 2,\mathbb{R}\right) }{U\left( 1\right) }\times \frac{%
SO\left( 6,n-2\right) }{SO\left( 6\right) \times SO\left( n-2\right) }%
,~n\geqslant 2~$ & $1+\text{min}\left( 6,n-2\right) ~$ & $\mathbb{R}\oplus
\mathbf{\Gamma }_{5,n-3}$ \\ \hline
\end{tabular}
\end{center}
\caption{\textit{Calabi-Vesentini} $d=4$ supergravity models. ``$mcs$''
stands for \textit{maximal compact subgroup} (with symmetric embedding). The
\textit{rank} of the scalar manifold, as well as the related \textit{%
reducible} Euclidean rank-$3$ Jordan algebra $J_{3}$ are also given (for
further elucidation and a recent treatment, see \textit{e.g.} \protect\cite
{Small-Orbits-Physics,Small-Orbits-Maths} and Refs. therein). The subscript
``$v $'' stands for \textit{``vertical''}, and it has been introduced in
order to distinguish the $S$-duality $SL_{v}\left( 2,\mathbb{R}\right) $
group from the ``horizontal'' symmetry group $SL_{h}\left( 2,\mathbb{R}%
\right) $ }
\end{table}

By defining
\begin{equation}
p^{2}\equiv p^{\Lambda }p^{\Sigma }\eta _{\Lambda \Sigma },~q^{2}\equiv
q_{\Lambda }q_{\Sigma }\eta ^{\Lambda \Sigma },~p\cdot q\equiv p^{\Lambda
}q_{\Lambda },
\end{equation}
where $\eta _{\Lambda \Sigma }=\eta ^{\Lambda \Sigma }$ is the
pseudo-Euclidean metric of $SO\left( m,n-2\right) $, the unique
algebraically-independent single-centered $G_{4}$-invariant polynomial $%
I_{4} $ (homogeneous of order $4$ in the fluxes) reads \cite{CY,DLR,CT}
\begin{equation}
I_{4}\left( \mathcal{Q}\right) \equiv p^{2}q^{2}-\left( p\cdot q\right) ^{2},
\label{I4-CV}
\end{equation}
and, by virtue of the CV covariant split (\ref{CV-split}), it can be
rewritten as:
\begin{equation}
I_{4}\left( \mathcal{Q}\right) =\frac{1}{2}\epsilon ^{\alpha \beta }\epsilon
^{\gamma \delta }\eta _{\Lambda \Xi }\eta _{\Sigma \Omega }\mathcal{Q}%
_{\alpha }^{\Lambda }\mathcal{Q}_{\beta }^{\Sigma }\mathcal{Q}_{\gamma
}^{\Xi }\mathcal{Q}_{\delta }^{\Omega }\equiv \frac{1}{2}\mathbb{K}_{\Lambda
\Sigma \Xi \Omega }^{\alpha \beta \gamma \delta }\mathcal{Q}_{\alpha
}^{\Lambda }\mathcal{Q}_{\beta }^{\Sigma }\mathcal{Q}_{\gamma }^{\Xi }%
\mathcal{Q}_{\delta }^{\Omega },  \label{red-red-red}
\end{equation}
where $\mathbb{K}_{\Lambda \Sigma \Xi \Omega }^{\alpha \beta \gamma \delta }$
is the $G_{4}$-invariant rank-$4$ completely symmetric $\mathbb{K}$-tensor $%
\mathbb{K}_{MNPQ}$ (see \textit{e.g.} \cite{Exc-Reds} and Refs. therein) of
the CV models, which enjoys the \textit{reducible} expression
\begin{equation}
\mathbb{K}_{\Lambda \Sigma \Xi \Omega }^{\alpha \beta \gamma \delta }\equiv
\frac{1}{6}\left[ \left( \epsilon ^{\alpha \beta }\epsilon ^{\gamma \delta
}+\epsilon ^{\alpha \delta }\epsilon ^{\beta \gamma }\right) \eta _{\Lambda
\Xi }\eta _{\Sigma \Omega }+\left( \epsilon ^{\alpha \beta }\epsilon
^{\delta \gamma }+\epsilon ^{\alpha \gamma }\epsilon ^{\delta \beta }\right)
\eta _{\Lambda \Omega }\eta _{\Sigma \Xi }+\left( \epsilon ^{\alpha \gamma
}\epsilon ^{\beta \delta }+\epsilon ^{\alpha \delta }\epsilon ^{\beta \gamma
}\right) \eta _{\Lambda \Sigma }\eta _{\Xi \Omega }\right]  \label{K-T}
\end{equation}
in term of the invariant structures $\epsilon ^{\alpha \beta }$ and $\eta
_{\Lambda \Xi }$ of $SL_{v}\left( 2,\mathbb{R}\right) $ and of $SO\left(
m,n-2\right) $, respectively.

We recall that the rank-$2$ antisymmetric $T$-tensor
\begin{equation}
T^{\Lambda \Sigma }\equiv p^{\Lambda }q^{\Sigma }-q^{\Lambda }p^{\Sigma }
\label{1-ctr}
\end{equation}
plays a key role in the classification of single-centered black hole (BH)
charge orbits in CV models (see \textit{e.g.} \cite
{Ferrara-Maldacena,ADF-fixed,CFMZ1,ADFT-FO-1}); furthermore, we anticipate
that $T^{\Lambda \Sigma }$ (\ref{1-ctr}) is the ``$1$-centered limit'' $%
1\equiv 2$ of the tensor $T_{\left( ab\right) }^{\left[ \Lambda \Sigma %
\right] }$ defined by the third of (\ref{T-Tensors}) further below, relevant
for the treatment of $2$-centered BH solutions. As it is well known, the
order-$4$ invariant $\mathcal{I}_{4}\left( \mathcal{Q}\right) $ (\ref{I4-CV}%
)-(\ref{red-red-red}) also enjoys a simple expression in terms of the tensor
$T^{\Lambda \Sigma }$ (\ref{1-ctr}):
\begin{equation}
I_{4}\left( \mathcal{Q}\right) =\frac{1}{2}T^{\Lambda \Sigma }T^{\Xi \Omega
}\eta _{\Lambda \Xi }\eta _{\Sigma \Omega }=-\frac{1}{2}\text{Tr}_{\eta
}\left( \mathbb{T}^{2}\right) ,
\end{equation}
where ``Tr$_{\eta }$'' denotes throughout the $\eta $-trace, namely the
trace in which the $SO\left( m,n-2\right) $ vector indices are consistently
raised and lowered by the $\eta $-structure.\medskip

From (\ref{G4-red}), the (``horizontal'' $\times $ $U$-duality) group of a $%
2 $-centered solution in $D=4$ CV models reads
\begin{equation}
\mathcal{G}_{p=2}\times G_{4}=SL_{h}\left( 2,\mathbb{R}\right) \times
SL_{v}\left( 2,\mathbb{R}\right) \times SO\left( m,n-2\right) \sim SO\left(
2,2\right) _{h}^{v}\times SO\left( m,n-2\right) ,  \label{1}
\end{equation}
where we recall that ``$h$'' and ``$v$'' respectively stand for
``horizontal'' and ``vertical''. In the $\mathcal{N}=2$ case, the number of
(Abelian) vector multiplets coupled to the gravity multiplet is $n_{V,%
\mathcal{N}=2}=n-1$. We will throughout consider $2$-centered $0$-brane (BH)
solutions, and thus the relevant representation of $\mathcal{G}_{p=2}\times
G_{4}$ in which the corresponding $2$-form field strengths fluxes $\mathcal{Q%
}_{a\alpha }^{\Lambda }$ sit is
\begin{equation}
\left( \mathbf{2},\mathbf{2},\mathbf{n}\right) ~\text{of~}SL_{h}\left( 2,%
\mathbb{R}\right) \times SL_{v}\left( 2,\mathbb{R}\right) \times SO\left(
2,n-2\right) ,  \label{2}
\end{equation}
which is thus amenable to a \textit{Gramian} treatment, as considered in
Sec. 8 of \cite{FMOSY-1}. In the following treatment, the $\mathbf{2}$ of $%
SL_{h}\left( 2,\mathbb{R}\right) $ is spanned by the Latin lowercase indices
$a=1,2$, the $\mathbf{2}$ of $SL_{v}\left( 2,\mathbb{R}\right) $ is spanned
by the Greek lowercase indices $\alpha =1,2$, and the vector $\mathbf{n}$ of
$SO\left( 2,n-2\right) $ is spanned by the uppercase Greek indices $\Lambda
=0,1,...,n-1$ (``$0$'' pertaining to the graviphotonic fluxes, as mentioned
above). If no further decomposition with respect to proper subgroups is
considered, maximal $G_{4}$-covariance is manifest, and, as stated, the
symplectic frame under consideration is usually dubbed CV \cite{CV} frame
\cite{CDFVP-1}.\smallskip

Aim of the present note is to give a complete treatment of $2$-centered $%
G_{4}$- (``duality'') and ($\mathcal{G}_{p=2}\times G_{4}$)- (dubbed
``horizontal'') invariant homogeneous polynomial structures in the CV
symplectic frame at order $2$, $4$ and $6$ in the fluxes, thus clarifying,
generalizing and completing the treatment given in \cite{FMOSY-1}, whose
notation and formul\ae\ we will often refer to (reporting some of them, for
ease of consultation). We will also briefly comment on the $2$-centered
``horizontal'' symmetry of supergravity models with pseudo-unitary $U$%
-duality groups, refining the analysis of \cite{MS-FMO-1}.\medskip

The plan of the note is as follows.

In Sec. \ref{p=2-CV} we analyze the duality- and ``horizontal''- invariant
two-centered homogeneous polynomials in Calabi-Vesentini $D=4$ supergravity
models, at order $2$ (Sec. \ref{Order-2}), $4$ (\ref{Order-4}) and $6$ (\ref
{Order-6}) in the fluxes $\mathcal{Q}$'s, which is enough to determine the
corresponding complete \textit{``minimal degree''} bases (see discussion in
Sec. \ref{Remarks-CV}).

Then, in Sec. \ref{Hor-U(n,m)} we study duality- and ``horizontal''-
invariant two-centered polynomials in $D=4$ symmetric supergravity models
with $U$-duality group $G_{4}$ given by the pseudo-unitary group $U\left(
r,s\right) $.

The final Sec. \ref{Conclusion} contains various remarks and observations,
concerning the CV models (Sec. \ref{Remarks-CV}) and models with $%
G_{4}=U\left( r,s\right) $ (Sec. \ref{Remarks-U(n,m)}). Moreover, in Sec.
\ref{GSKG}, by suitably generalizing the notion of groups `\textit{`of type $%
E_{7}$"} \cite{CFMZ1,Brown-Groups-of-type-E7,FMY-FD-1}, we comment on their
relation to special K\"{a}hler geometry.

%the physical meaning and applications of the invariants under consideration
%(Sec. \ref{Remarks-Physics}).

\section{\label{p=2-CV}Calabi-Vesentini Flux Tensors and Invariants}

\subsection{\label{Order-2}Order $2$}

We start and consider the rank-$2$ tensor product
\begin{equation}
\mathcal{Q}_{a\alpha }^{\Lambda }\mathcal{Q}_{b\beta }^{\Sigma }\equiv
\left( \underset{a}{\mathbf{2}},~\underset{\alpha }{\mathbf{2}},~\underset{%
\Lambda }{\mathbf{n}}\right) \times _{s}\left( \underset{b}{\mathbf{2}},~%
\underset{\beta }{\mathbf{2}},~\underset{\Sigma }{\mathbf{n}}\right) =\left(
\mathbf{3}_{s}+\mathbf{1}_{a},~\mathbf{3}_{s}+\mathbf{1}_{a},~\left[ \frac{%
\mathbf{n}\left( \mathbf{n+1}\right) }{\mathbf{2}}\right] _{s}+\left[ \frac{%
\mathbf{n}\left( \mathbf{n-1}\right) }{\mathbf{2}}\right] _{a}\right) _{s},
\label{3}
\end{equation}
where ``$s$'' and ``$a$'' denote the symmetric and antisymmetric parts
throughout. \textit{\c{C}a va sans dire}, the obvious symmetry of $\mathcal{Q%
}_{a\alpha }^{\Lambda }\mathcal{Q}_{b\beta }^{\Sigma }$ under the exchange $%
a\alpha \Lambda \leftrightarrow b\beta \Sigma $ restricts the analysis to
the \textit{symmetric part} of such a tensor product. Since in
(pseudo-)orthogonal groups the symmetric rank-$2$ repr. can be further
irreducibly split in traceless $\mathbf{S}_{0}$ and trace $\mathbf{1}$
irreps. (with the naught denoting $\eta $-tracelessness in $SO\left(
2,n-2\right) $ throughout; see Footnote 4 of \cite{FMOSY-1}):
\begin{equation}
\frac{\mathbf{n}\left( \mathbf{n+1}\right) }{\mathbf{2}}=\mathbf{S}_{0}+%
\mathbf{1},  \label{4}
\end{equation}
(\ref{3}) can be further elaborated as:
\begin{equation}
\mathcal{Q}_{a\alpha }^{\Lambda }\mathcal{Q}_{b\beta }^{\Sigma }=\left(
\mathbf{3},~\mathbf{3},~\mathbf{S}_{0}\right) +\left( \mathbf{3},~\mathbf{3}%
,~\mathbf{1}\right) +\left( \mathbf{3},~\mathbf{1},~\mathbf{Adj}\right)
+\left( \mathbf{1},~\mathbf{3},~\mathbf{Adj}\right) +\left( \mathbf{1},~%
\mathbf{1},~\mathbf{S}_{0}\right) +\left( \mathbf{1},~\mathbf{1},~\mathbf{1}%
\right) ,  \label{5}
\end{equation}
where the following notation for $SO\left( 2,n-2\right) $ irreps. has been
introduced:
\begin{eqnarray}
\mathbf{S}_{0} &\equiv &\frac{\mathbf{n}\left( \mathbf{n+1}\right) }{\mathbf{%
2}}-\mathbf{1}~\text{(}\eta \text{-traceless~rank-}2\text{~symmetric)}; \\
\mathbf{Adj} &\equiv &\frac{\mathbf{n}\left( \mathbf{n-1}\right) }{\mathbf{2}%
}~\text{(rank-}2\text{~antisymmetric,~\textit{i.e.}~adjoint)}.
\end{eqnarray}
The total real dimension of $\mathcal{Q}_{a\alpha }^{\Lambda }\mathcal{Q}%
_{b\beta }^{\Sigma }$ is $2n\left( 4n+1\right) $. Thence, one can assign to
each irreps. a tensor structure (with ``$\#$'' denoting the corresponding
dimension):
\begin{equation}
\begin{array}{lll}
\begin{array}{l}
\left( \mathbf{3},~\mathbf{3},~\mathbf{S}_{0}\right) : \\
~
\end{array}
&
\begin{array}{l}
T_{\left( ab\right) ~\left( \alpha \beta \right) }^{0~\left( \Lambda \Sigma
\right) }\equiv T_{\left( ab\right) ~\left( \alpha \beta \right) }^{\left(
\Lambda \Sigma \right) }-\frac{1}{n}\eta ^{\Lambda \Sigma }\text{Tr}_{\eta
}\left( T_{\left( ab\right) ~\left( \alpha \beta \right) }\right) , \\
~
\end{array}
&
\begin{array}{l}
\#=\frac{9}{2}n\left( n+1\right) -1; \\
~
\end{array}
\\
\begin{array}{l}
\left( \mathbf{3},~\mathbf{3},~\mathbf{1}\right) : \\
~
\end{array}
&
\begin{array}{l}
\mathbf{T}_{\left( ab\right) ~\left( \alpha \beta \right) }^{\left( \Lambda
\Sigma \right) }\equiv \frac{1}{n}\eta ^{\Lambda \Sigma }\text{Tr}_{\eta
}\left( T_{\left( ab\right) ~\left( \alpha \beta \right) }\right) , \\
~
\end{array}
&
\begin{array}{l}
\#=1; \\
~
\end{array}
\\
\begin{array}{l}
\left( \mathbf{3},~\mathbf{1},~\mathbf{Adj}\right) : \\
~
\end{array}
&
\begin{array}{l}
T_{\left( ab\right) ~\left[ \alpha \beta \right] }^{\left[ \Lambda \Sigma
\right] }\Rightarrow T_{\left( ab\right) }^{\left[ \Lambda \Sigma \right]
}\equiv \epsilon ^{\alpha \beta }T_{\left( ab\right) ~\left[ \alpha \beta
\right] }^{\left[ \Lambda \Sigma \right] }, \\
~
\end{array}
&
\begin{array}{l}
\#=\frac{3}{2}n(n+1); \\
~
\end{array}
\\
\begin{array}{l}
\left( \mathbf{1},~\mathbf{3},~\mathbf{Adj}\right) : \\
~
\end{array}
&
\begin{array}{l}
T_{\left[ ab\right] ~\left( \alpha \beta \right) }^{\left[ \Lambda \Sigma
\right] }\Rightarrow T_{\left( \alpha \beta \right) }^{\left[ \Lambda \Sigma
\right] }\equiv \epsilon ^{ab}T_{\left( ab\right) ~\left[ \alpha \beta
\right] }^{\left[ \Lambda \Sigma \right] }, \\
~
\end{array}
&
\begin{array}{l}
\#=\frac{3}{2}n(n+1); \\
~
\end{array}
\\
\begin{array}{c}
\left( \mathbf{1},~\mathbf{1},~\mathbf{S}_{0}+\mathbf{1}\right) : \\
~
\end{array}
&
\begin{array}{c}
T_{\left[ ab\right] ~\left[ \alpha \beta \right] }^{\left( \Lambda \Sigma
\right) }\Rightarrow \left\{
\begin{array}{l}
T_{\left[ \alpha \beta \right] }^{\left( \Lambda \Sigma \right) }\equiv
\epsilon ^{ab}T_{\left[ ab\right] ~\left[ \alpha \beta \right] }^{\left(
\Lambda \Sigma \right) }; \\
\\
T_{\left[ ab\right] }^{\left( \Lambda \Sigma \right) }\equiv \epsilon
^{\alpha \beta }T_{\left[ ab\right] ~\left[ \alpha \beta \right] }^{\left(
\Lambda \Sigma \right) }; \\
\\
T^{\left( \Lambda \Sigma \right) }\equiv \epsilon ^{ab}\epsilon ^{\alpha
\beta }T_{\left[ ab\right] ~\left[ \alpha \beta \right] }^{\left( \Lambda
\Sigma \right) }.
\end{array}
\right. \\
~
\end{array}
&
\begin{array}{l}
\#=\frac{1}{2}n\left( n+1\right) ; \\
~\text{(each~of~them)}
\end{array}
\end{array}
\label{T-Tensors}
\end{equation}
Below, we will also use the further irreducibly split $\left( \mathbf{1},~%
\mathbf{1},~\mathbf{S}_{0}+\mathbf{1}\right) $ (as (\ref{4})), reading:
\begin{equation}
\begin{array}{ccc}
\begin{array}{l}
\left( \mathbf{1},~\mathbf{1},~\mathbf{S}_{0}\right) : \\
~
\end{array}
&
\begin{array}{l}
T_{\left[ ab\right] ~\left[ \alpha \beta \right] }^{0~\left( \Lambda \Sigma
\right) }\equiv T_{\left[ ab\right] ~\left[ \alpha \beta \right] }^{\left(
\Lambda \Sigma \right) }-\frac{1}{n}\eta ^{\Lambda \Sigma }\text{Tr}_{\eta
}\left( T_{\left[ ab\right] ~\left[ \alpha \beta \right] }\right) , \\
~
\end{array}
&
\begin{array}{l}
\#=\frac{1}{2}n\left( n+1\right) -1; \\
~
\end{array}
\\
\begin{array}{l}
\left( \mathbf{1},~\mathbf{1},~\mathbf{1}\right) : \\
~
\end{array}
&
\begin{array}{l}
\mathbf{T}_{\left[ ab\right] ~\left[ \alpha \beta \right] }^{\left( \Lambda
\Sigma \right) }\equiv \frac{1}{n}\eta ^{\Lambda \Sigma }\text{Tr}_{\eta
}\left( T_{\left[ ab\right] ~\left[ \alpha \beta \right] }\right) , \\
~
\end{array}
&
\begin{array}{l}
\#=1. \\
~
\end{array}
\end{array}
\end{equation}
Explicit expressions in terms of the flux vector $\mathcal{Q}_{a\alpha
}^{\Lambda }$ read\footnote{%
In order to make contact with the notation of \cite{FMOSY-1}, we observe
that $T_{\left( ab\right) }^{\left[ \Lambda \Sigma \right] }$ (\ref{def-1};
also see (\ref{mn-T'})) and $T^{\left( \Lambda \Sigma \right) }$ (\ref{def-2}%
; also see (\ref{mn-T''})) respectively correspond to $\mathbb{T}\equiv
\left( \mathbb{T}_{11}\equiv \mathbb{T}_{1},\mathbb{T}_{12},\mathbb{T}%
_{22}\equiv \mathbb{T}_{2}\right) $ and $2\mathbb{T}_{a}^{\Lambda \Sigma }$,
in turn given by Eqs. (3.3)-(3.6) resp. (4.2) of \cite{FMOSY-1}.}:
\begin{eqnarray}
T_{\left( ab\right) ~\left( \alpha \beta \right) }^{\left( \Lambda \Sigma
\right) } &=&\frac{1}{4}\left( \mathcal{Q}_{a\alpha }^{\Lambda }\mathcal{Q}%
_{b\beta }^{\Sigma }+\mathcal{Q}_{a\alpha }^{\Sigma }\mathcal{Q}_{b\beta
}^{\Lambda }+\mathcal{Q}_{a\beta }^{\Lambda }\mathcal{Q}_{b\alpha }^{\Sigma
}+\mathcal{Q}_{a\beta }^{\Sigma }\mathcal{Q}_{b\alpha }^{\Lambda }\right) ;
\\
T_{\left( ab\right) }^{\left[ \Lambda \Sigma \right] } &=&\frac{1}{2}\left(
\mathcal{Q}_{a\alpha }^{\Lambda }\mathcal{Q}_{b\beta }^{\Sigma }-\mathcal{Q}%
_{a\alpha }^{\Sigma }\mathcal{Q}_{b\beta }^{\Lambda }\right) \epsilon
^{\alpha \beta };  \label{def-1} \\
T_{\left( \alpha \beta \right) }^{\left[ \Lambda \Sigma \right] } &=&\frac{1%
}{2}\left( \mathcal{Q}_{a\alpha }^{\Lambda }\mathcal{Q}_{b\beta }^{\Sigma }-%
\mathcal{Q}_{a\beta }^{\Sigma }\mathcal{Q}_{b\alpha }^{\Lambda }\right)
\epsilon ^{ab}; \\
T^{\left( \Lambda \Sigma \right) } &=&\mathcal{Q}_{a\alpha }^{\Lambda }%
\mathcal{Q}_{b\beta }^{\Sigma }\epsilon ^{ab}\epsilon ^{\alpha \beta }.
\label{def-2}
\end{eqnarray}
We will also make use of the following $SO\left( 2,n-2\right) $-matrix
notations:
\begin{eqnarray}
T_{\left( ab\right) ~\left( \alpha \beta \right) } &\equiv &T_{\left(
ab\right) ~\left( \alpha \beta \right) }^{\left( \Lambda \Sigma \right) };
\label{mn-T} \\
T_{\left( ab\right) }^{\prime } &\equiv &T_{\left( ab\right) }^{\left[
\Lambda \Sigma \right] };  \label{mn-T'} \\
T_{\left[ ab\right] ~\left[ \alpha \beta \right] }^{\prime \prime } &\equiv
&T_{\left[ ab\right] ~\left[ \alpha \beta \right] }^{\left( \Lambda \Sigma
\right) },  \label{mn-T''}
\end{eqnarray}
and analogous ones for the $\epsilon $-traces.

By further taking (half of) the $\eta $-trace of $T^{\left( \Lambda \Sigma
\right) }$, one obtains the \textit{symplectic product} $\mathcal{W}$ of the
two charge vectors $\mathcal{Q}_{1\alpha }^{\Lambda }$ and $\mathcal{Q}%
_{2\alpha }^{\Lambda }$ (\textit{cfr.} \textit{e.g.} (4.12) of \cite{FMOSY-1}%
, as well as (3.9) of \cite{ADFMT-1}):
\begin{equation}
\mathcal{W}=\frac{1}{2}\eta _{\Lambda \Sigma }T^{\left( \Lambda \Sigma
\right) }=\frac{1}{2}\epsilon ^{ab}\epsilon ^{\alpha \beta }\eta _{\Lambda
\Sigma }T_{\left[ ab\right] ~\left[ \alpha \beta \right] }^{\left( \Lambda
\Sigma \right) }=\frac{1}{2}\text{Tr}_{\eta }\left( T^{\prime \prime
}\right) ,  \label{15}
\end{equation}
which is evidently ``horizontal'' (\textit{i.e.} $\left( \mathcal{G}%
_{p=2}\times G_{4}\right) $-) invariant (actually, as yielded by the
analysis of \cite{FMOSY-1} and \cite{ADFMT-1}, the unique
``horizontal''-invariant polynomial at order $2$ in the fluxes).\smallskip\
Clearly, (\ref{15}) is a specification for CV models of the general formula (%
\textit{cfr.} \textit{e.g.} Eq. (3.9) of \cite{ADFMT-1}, and Sec. 3 therein
for notation)
\begin{equation}
\mathcal{W}\equiv \frac{1}{2}\mathbb{C}_{MN}\epsilon ^{ab}\mathcal{Q}_{a}^{M}%
\mathcal{Q}_{b}^{N},  \label{15-bis}
\end{equation}
where $\mathbb{C}_{MN}$ is the symplectic-invariant metric defined in (\ref
{C-structure}) below.

The $\epsilon $-traced tensors in (\ref{T-Tensors}) have been introduced in
order to develop the subsequent treatment. Indeed, due to the very structure
of (\ref{4}), the irreducible splitting (\ref{4}) is not relevant in order
to classify and relate duality- and ``horizontal''- invariant homogeneous
polynomials in the BH fluxes (see the treatment of Secs. \ref{Order-4} and
\ref{Order-6}).

\subsection{\label{Order-4}Order $4$}

Since there is no duality- nor ``horizontal''- invariant polynomial
structure at order $3$, next we proceed to analyze the order $4$ in the
fluxes of the $2$-centered BH solution in the framework under consideration.
By exploiting the associativity of the (ir)reps.' tensor product, in each of
the $SL_{h}\left( 2,\mathbb{R}\right) $- and $SL_{v}\left( 2,\mathbb{R}%
\right) $- sectors one gets
\begin{equation}
\mathbf{2}\times \mathbf{2}\times \mathbf{2}\times \mathbf{2}=\mathbf{3}%
\times \mathbf{3}+2\cdot \left( \mathbf{3}\times \mathbf{1}\right) +\mathbf{1%
}\times \mathbf{1=5+}3\cdot \mathbf{3}+2\cdot \mathbf{1},  \label{16}
\end{equation}
whereas in the $SO\left( 2,n-2\right) $ sector, recalling that
\begin{eqnarray}
\mathbf{S}_{0}\times \mathbf{S}_{0} &=&\mathbf{1}_{s}+\mathbf{Adj}_{a}+%
\mathbf{S}_{0,s}+...;  \label{17} \\
\mathbf{Adj}\times \mathbf{Adj} &=&\mathbf{1}_{s}+\mathbf{Adj}_{a}+\mathbf{S}%
_{0,s}+...;  \label{18} \\
\mathbf{Adj}\times \mathbf{S}_{0} &=&\mathbf{Adj}+\mathbf{S}_{0}+...,
\label{19}
\end{eqnarray}
it holds
\begin{equation}
\mathbf{n}\times \mathbf{n}\times \mathbf{n}\times \mathbf{n}=\left( \mathbf{%
S}_{0}+\mathbf{1}+\mathbf{Adj}\right) \times \left( \mathbf{S}_{0}+\mathbf{1}%
+\mathbf{Adj}\right) =3\cdot \mathbf{1}+6\cdot \mathbf{Adj}+6\cdot \mathbf{S}%
_{0}+...~.  \label{20}
\end{equation}

Thus, the duality- or ``horizontal''- invariant homogeneous polynomials at
order $4$ in the fluxes arise from the following tensor products:

\begin{enumerate}
\item
\begin{equation}
\left( \mathbf{3},\mathbf{3},\mathbf{S}_{0}\right) \times \left( \mathbf{3},%
\mathbf{3},\mathbf{S}_{0}\right) =\underset{G_{4}\text{-inv.}}{\left(
\mathbf{3}\times \mathbf{3},\mathbf{1},\mathbf{1}\right) }+...  \label{21}
\end{equation}
Since $\mathbf{1}\notin \mathbf{S}_{2}\times \mathbf{Adj}$, there are no
other sources of duality-invariant polynomials involving tensor products of $%
\left( \mathbf{3},\mathbf{3},\mathbf{S}_{0}\right) $. By using the $%
SO(2,n-2) $-matrix notation (\ref{mn-T}), the order-$4$ $G_{4}$-invariant
polynomial from (\ref{21}) can reducibly be rewritten as the $SL_{h}\left( 2,%
\mathbb{R}\right) $-bi-triplet
\begin{eqnarray}
\mathcal{J}_{\left( ab\right) \left( cd\right) }^{0} &\equiv &-T_{\left(
ab\right) ~\left( \alpha \beta \right) }^{0~\left( \Lambda \Sigma \right)
}T_{\left( cd\right) ~\left( \gamma \delta \right) }^{0~\left( \Xi \Delta
\right) }\eta _{\Lambda \Xi }\eta _{\Sigma \Delta }\epsilon ^{\alpha \gamma
}\epsilon ^{\beta \delta }  \notag \\
&=&-\left[ T_{\left( ab\right) ~\left( \alpha \beta \right) }^{\left(
\Lambda \Sigma \right) }-\frac{1}{n}\eta ^{\Lambda \Sigma }\text{Tr}_{\eta
}\left( T_{\left( ab\right) ~\left( \alpha \beta \right) }\right) \right] %
\left[ T_{\left( cd\right) ~\left( \gamma \delta \right) }^{\left( \Xi
\Delta \right) }-\frac{1}{n}\eta ^{\Xi \Delta }\text{Tr}_{\eta }\left(
T_{\left( cd\right) ~\left( \gamma \delta \right) }\right) \right] \eta
_{\Lambda \Xi }\eta _{\Sigma \Delta }\epsilon ^{\alpha \gamma }\epsilon
^{\beta \delta }  \notag \\
&=&-T_{\left( ab\right) ~\left( \alpha \beta \right) }^{\left( \Lambda
\Sigma \right) }T_{\left( cd\right) ~\left( \gamma \delta \right) }^{\left(
\Xi \Delta \right) }\eta _{\Lambda \Xi }\eta _{\Sigma \Delta }\epsilon
^{\alpha \gamma }\epsilon ^{\beta \delta }+\frac{1}{n}\text{Tr}_{\eta
}\left( T_{\left( ab\right) ~\left( \alpha \beta \right) }\right) \text{Tr}%
_{\eta }\left( T_{\left( cd\right) ~\left( \gamma \delta \right) }\right)
\epsilon ^{\alpha \gamma }\epsilon ^{\beta \delta }.  \label{22}
\end{eqnarray}
It is convenient to introduce the following tensors (see also point 2
below):
\begin{eqnarray}
\mathcal{J}_{\left( ab\right) \left( cd\right) } &\equiv &-T_{\left(
ab\right) ~\left( \alpha \beta \right) }^{\left( \Lambda \Sigma \right)
}T_{\left( cd\right) ~\left( \gamma \delta \right) }^{\left( \Xi \Delta
\right) }\eta _{\Lambda \Xi }\eta _{\Sigma \Delta }\epsilon ^{\alpha \gamma
}\epsilon ^{\beta \delta };  \label{22-2} \\
\mathbf{J}_{\left( ab\right) \left( cd\right) } &\equiv &\mathbf{T}_{\left(
ab\right) ~\left( \alpha \beta \right) }^{\left( \Lambda \Sigma \right) }%
\mathbf{T}_{\left( cd\right) ~\left( \gamma \delta \right) }^{\left( \Xi
\Delta \right) }\eta _{\Lambda \Xi }\eta _{\Sigma \Delta }\epsilon ^{\alpha
\gamma }\epsilon ^{\beta \delta }=\frac{1}{n}\text{Tr}_{\eta }\left(
T_{\left( ab\right) ~\left( \alpha \beta \right) }\right) \text{Tr}_{\eta
}\left( T_{\left( cd\right) ~\left( \gamma \delta \right) }\right) \epsilon
^{\alpha \gamma }\epsilon ^{\beta \delta },  \notag \\
&&  \label{22-3}
\end{eqnarray}
such that Eq. (\ref{22}) can be rewritten as
\begin{equation}
\mathcal{J}_{\left( ab\right) \left( cd\right) }^{0}=\mathcal{J}_{\left(
ab\right) \left( cd\right) }+\mathbf{J}_{\left( ab\right) \left( cd\right) }.
\end{equation}
Then, $\mathcal{J}_{\left( ab\right) \left( cd\right) }$ (\ref{22-2}) can be
$SL_{h}\left( 2,\mathbb{R}\right) $-irreducibly decomposed as
\begin{equation}
\underset{\mathbf{3}\times \mathbf{3}}{\mathcal{J}_{\left( ab\right) \left(
cd\right) }}=\underset{\mathbf{5}_{s}^{0}+\mathbf{1}_{s}}{-\text{Tr}_{\eta
}\left( T_{(\left( ab\right) }T_{\left( cd\right) )}\right) }\underset{%
\mathbf{3}_{a}}{-\text{Tr}_{\eta }\left( T_{[\left( ab\right) }T_{\left(
cd\right) ]}\right) }.
\end{equation}
By making use of the cyclic property of Tr$_{\eta }$ and of the
distributivity of the sum with respect to it, one obtains
\begin{equation}
\mathbf{3}_{a}\equiv -\text{Tr}_{\eta }\left( T_{[\left( ab\right)
}T_{\left( cd\right) ]}\right) =0,
\end{equation}
and therefore $\mathcal{J}_{\left( ab\right) \left( cd\right) }$ can be
rewritten as
\begin{equation}
\mathcal{J}_{\left( ab\right) \left( cd\right) }=-\text{Tr}_{\eta }\left(
T_{\left( ab\right) }T_{\left( cd\right) }\right) =-\text{Tr}_{\eta }\left(
T_{(\left( ab\right) }T_{\left( cd\right) )}\right) =-\frac{1}{3}\text{Tr}%
_{\eta }\left( T_{\left( ab\right) }T_{\left( cd\right) }+T_{\left(
ac\right) }T_{\left( bd\right) }+T_{\left( ad\right) }T_{\left( bc\right)
}\right) ,
\end{equation}
with
\begin{eqnarray}
\mathbf{5}_{s}^{0} &\equiv &\mathcal{J}_{\left( abcd\right) }\equiv \mathcal{%
J}_{\left( ab\right) \left( cd\right) }-\frac{1}{3}\left( \mathcal{X}-\frac{5%
}{2}\mathcal{W}^{2}\right) \epsilon _{a\left( c\right| }\epsilon _{b\left|
d\right) };  \label{r-0} \\
\mathbf{1}_{s} &\equiv &\mathcal{J}_{\left( ab\right) \left( cd\right) }-%
\mathcal{J}_{\left( abcd\right) }=\frac{1}{3}\epsilon ^{a^{\prime }c^{\prime
}}\epsilon ^{b^{\prime }d^{\prime }}\text{Tr}_{\eta }\left( T_{\left(
a^{\prime }b^{\prime }\right) }T_{\left( c^{\prime }d^{\prime }\right)
}+T_{\left( a^{\prime }d^{\prime }\right) }T_{\left( b^{\prime }c^{\prime
}\right) }\right) \epsilon _{a\left( c\right| }\epsilon _{b\left| d\right) }
\notag \\
&=&\frac{1}{3}\left( \mathcal{X}-\frac{5}{2}\mathcal{W}^{2}\right) \epsilon
_{a\left( c\right| }\epsilon _{b\left| d\right) }.  \label{r-1}
\end{eqnarray}
$\mathcal{X}$ is the order-$4$ ``horizontal'' invariant homogeneous
polynomial defined by (4.13) of \cite{FMOSY-1}, which we report here, in the
current notation (recall (\ref{mn-T})-(\ref{mn-T''}), as well as Footnote
3):
\begin{equation}
\mathcal{X}\equiv -\text{Tr}_{\eta }\left( T_{11}^{\prime }T_{22}^{\prime
}\right) +\text{Tr}_{\eta }\left( T_{12}^{\prime 2}\right) -\frac{1}{8}\text{%
Tr}_{\eta }^{2}\left( T^{\prime \prime }\right) .  \label{Chi-def}
\end{equation}
In order to get an ``horizontal'' invariant polynomial homogeneous of order $%
4$ in the fluxes, one has \textit{e.g.} to $\epsilon $-trace both sides of (%
\ref{r-1}), obtaining (as a consequence of the $\epsilon $-tracelessness of $%
\mathcal{J}_{\left( abcd\right) }$ (\ref{r-0})):
\begin{equation}
\epsilon ^{ac}\epsilon ^{bd}\mathcal{J}_{\left( ab\right) \left( cd\right) }=%
\frac{1}{3}\left( \mathcal{X}-\frac{5}{2}\mathcal{W}^{2}\right) \epsilon
^{ac}\epsilon ^{bd}\epsilon _{a\left( c\right| }\epsilon _{b\left| d\right)
}=\mathcal{X}-\frac{5}{2}\mathcal{W}^{2}.
\end{equation}

\item
\begin{equation}
\left( \mathbf{3},\mathbf{3},\mathbf{1}\right) \times \left( \mathbf{3},%
\mathbf{3},\mathbf{1}\right) =\underset{G_{4}\text{-inv.}}{\left( \mathbf{3}%
\times \mathbf{3},\mathbf{1},\mathbf{1}\right) }+...  \label{23}
\end{equation}
There are no other sources of duality-invariant polynomials involving tensor
products of $\left( \mathbf{3},\mathbf{3},\mathbf{1}\right) $. The order-$4$
$G_{4}$-invariant polynomial from (\ref{23}) can reducibly be rewritten as
the $SL_{h}\left( 2,\mathbb{R}\right) $-bi-triplet $\mathbf{J}_{\left(
ab\right) \left( cd\right) }$ defined by (\ref{22-3}), which enjoys a
decomposition analogous to the one of $\mathcal{J}_{\left( ab\right) \left(
cd\right) }$.

\item
\begin{equation}
\left( \mathbf{3},\mathbf{1},\mathbf{Adj}\right) \times \left( \mathbf{3},%
\mathbf{1},\mathbf{Adj}\right) =\underset{G_{4}\text{-inv.}}{\left( \mathbf{3%
}\times \mathbf{3},\mathbf{1},\mathbf{1}\right) }+...  \label{24}
\end{equation}
Since $\mathbf{1}\notin \mathbf{S}_{2}\times \mathbf{Adj}$, there are no
other sources of duality-invariant polynomials involving tensor products of $%
\left( \mathbf{3},\mathbf{1},\mathbf{Adj}\right) $. The order-$4$ $G_{4}$%
-invariant polynomial from (\ref{24}) can reducibly be rewritten as the $%
SL_{h}\left( 2,\mathbb{R}\right) $-bi-triplet
\begin{equation}
T_{\left( ab\right) ~\left[ \alpha \beta \right] }^{\left[ \Lambda \Sigma %
\right] }T_{\left( cd\right) ~\left[ \gamma \delta \right] }^{\left[ \Xi
\Delta \right] }\eta _{\Lambda \Xi }\eta _{\Sigma \Delta }.  \label{25}
\end{equation}
However, without any loss of generality, one can instead consider (half of)
the tensor product of the corresponding $SL_{v}\left( 2,\mathbb{R}\right) $ $%
\epsilon $-traces (which is also the unique independent manifestly $%
SL_{v}\left( 2,\mathbb{R}\right) $-invariant combination); by using the $%
SO(2,n-2)$-matrix notation (\ref{mn-T'}), one obtains the following $%
SL_{h}\left( 2,\mathbb{R}\right) $-bi-triplet
\begin{eqnarray}
I_{\left( ab\right) \left( cd\right) } &\equiv &\frac{1}{2}T_{\left(
ab\right) }^{\left[ \Lambda \Sigma \right] }T_{\left( cd\right) }^{\left[
\Xi \Delta \right] }\eta _{\Lambda \Xi }\eta _{\Sigma \Delta }=\frac{1}{2}%
\epsilon ^{\alpha \beta }\epsilon ^{\gamma \delta }T_{\left( ab\right) ~%
\left[ \alpha \beta \right] }^{\left[ \Lambda \Sigma \right] }T_{\left(
cd\right) ~\left[ \gamma \delta \right] }^{\left[ \Xi \Delta \right] }\eta
_{\Lambda \Xi }\eta _{\Sigma \Delta }  \notag \\
&=&\frac{1}{2}T_{\left[ \Lambda \Sigma \right] \left( ab\right) }T_{\left(
cd\right) }^{\left[ \Lambda \Sigma \right] }=-\frac{1}{2}\text{Tr}_{\eta
}\left( T_{\left( ab\right) }^{\prime }T_{\left( cd\right) }^{\prime
}\right) .  \label{26}
\end{eqnarray}
Then, $I_{\left( ab\right) \left( cd\right) }$ can be $SL_{h}\left( 2,%
\mathbb{R}\right) $-irreducibly decomposed as
\begin{equation}
\underset{\mathbf{3}\times \mathbf{3}}{I_{\left( ab\right) \left( cd\right) }%
}=\underset{\mathbf{5}_{s}^{0}+\mathbf{1}_{s}}{-\frac{1}{2}\text{Tr}_{\eta
}\left( T_{(\left( ab\right) }^{\prime }T_{\left( cd\right) )}^{\prime
}\right) }\underset{\mathbf{3}_{a}}{-\frac{1}{2}\text{Tr}_{\eta }\left(
T_{[\left( ab\right) }^{\prime }T_{\left( cd\right) ]}^{\prime }\right) }.
\label{27}
\end{equation}
Since
\begin{equation}
\mathbf{3}_{a}\equiv -\frac{1}{2}\text{Tr}_{\eta }\left( T_{[\left(
ab\right) }^{\prime }T_{\left( cd\right) ]}^{\prime }\right) =0,  \label{30}
\end{equation}
$I_{\left( ab\right) \left( cd\right) }$ can be rewritten as
\begin{eqnarray}
I_{\left( ab\right) \left( cd\right) } &=&-\frac{1}{2}\text{Tr}_{\eta
}\left( T_{\left( ab\right) }^{\prime }T_{\left( cd\right) }^{\prime
}\right) =-\frac{1}{2}\text{Tr}_{\eta }\left( T_{(\left( ab\right) }^{\prime
}T_{\left( cd\right) )}^{\prime }\right)  \notag \\
&=&-\frac{1}{6}\text{Tr}_{\eta }\left( T_{\left( ab\right) }^{\prime
}T_{\left( cd\right) }^{\prime }+T_{\left( ac\right) }^{\prime }T_{\left(
bd\right) }^{\prime }+T_{\left( ad\right) }^{\prime }T_{\left( bc\right)
}^{\prime }\right) ,  \label{28-pre}
\end{eqnarray}
with
\begin{eqnarray}
\mathbf{5}_{s}^{0} &\equiv &I_{\left( abcd\right) }\equiv I_{\left(
ab\right) \left( cd\right) }-\frac{1}{3}\left( \mathcal{X}+\frac{1}{2}%
\mathcal{W}^{2}\right) \epsilon _{a\left( c\right| }\epsilon _{b\left|
d\right) };  \label{28} \\
\mathbf{1}_{s} &\equiv &I_{\left( ab\right) \left( cd\right) }-I_{\left(
abcd\right) }=\frac{1}{3}\epsilon ^{a^{\prime }c^{\prime }}\epsilon
^{b^{\prime }d^{\prime }}\text{Tr}_{\eta }\left( T_{\left( a^{\prime
}b^{\prime }\right) }^{\prime }T_{\left( c^{\prime }d^{\prime }\right)
}^{\prime }+T_{\left( a^{\prime }d^{\prime }\right) }^{\prime }T_{\left(
b^{\prime }c^{\prime }\right) }^{\prime }\right) \epsilon _{a\left( c\right|
}\epsilon _{b\left| d\right) }  \notag \\
&=&\frac{1}{3}\left( \mathcal{X}+\frac{1}{2}\mathcal{W}^{2}\right) \epsilon
_{a\left( c\right| }\epsilon _{b\left| d\right) }.  \label{29}
\end{eqnarray}
$I_{\left( abcd\right) }$ is the so-called \textit{Dixmier tensor} \cite
{Dixmier} (or better its ``two-centered analogue''), introduced in
supergravity in \cite{FMOSY-1,ADFMT-1,Small-1,FMY-FD-1}, and generally
related to the $\mathbb{K}$-tensor $\mathbb{K}_{\left( MNPQ\right) }$ of $%
G_{4}$ (\cite{Brown-Groups-of-type-E7}; see also \cite{Exc-Reds} and Refs.
therein) as follows\footnote{%
For a discussion of the differences between CV (\textit{i.e.} \textit{%
reducible} symmetric) and \textit{irreducible} symmetric $D=4$ supergravity
models, see \cite{ADFMT-1} and \cite{Small-1}.}:
\begin{equation}
I_{\left( abcd\right) }\equiv \frac{1}{2}\mathbb{K}_{MNPQ}\mathcal{Q}_{a}^{M}%
\mathcal{Q}_{b}^{N}\mathcal{Q}_{c}^{P}\mathcal{Q}_{d}^{Q}.
\label{Dixmier-tensor-def}
\end{equation}
In order to get an ``horizontal'' invariant polynomial homogeneous of order $%
4$ in the fluxes, one has then to $\epsilon $-trace the unique $\epsilon $%
-traceful quantity out of (\ref{28})-(\ref{29}), namely $\mathbf{1}_{s}$; by
also recalling Eq. (4.13) of \cite{FMOSY-1}, the following result
(consequence of the $\epsilon $-tracelessness of $I_{\left( abcd\right) }$ (%
\ref{28})) is achieved:
\begin{eqnarray}
I_{\left( ab\right) \left( cd\right) }\epsilon ^{ac}\epsilon ^{bd} &=&\frac{1%
}{2}T_{\left( ab\right) }^{\left[ \Lambda \Sigma \right] }T_{\left(
cd\right) }^{\left[ \Xi \Delta \right] }\eta _{\Lambda \Xi }\eta _{\Sigma
\Delta }\epsilon ^{ac}\epsilon ^{bd}=\frac{1}{2}\epsilon ^{\alpha \beta
}\epsilon ^{\gamma \delta }\epsilon ^{ac}\epsilon ^{bd}T_{\left( ab\right) ~%
\left[ \alpha \beta \right] }^{\left[ \Lambda \Sigma \right] }T_{\left(
cd\right) ~\left[ \gamma \delta \right] }^{\left[ \Xi \Delta \right] }\eta
_{\Lambda \Xi }\eta _{\Sigma \Delta }  \notag \\
&=&-\frac{1}{2}\epsilon ^{ac}\epsilon ^{bd}\text{Tr}_{\eta }\left( T_{\left(
ab\right) }^{\prime }T_{\left( cd\right) }^{\prime }\right) =\frac{1}{3}%
\left( \mathcal{X}+\frac{1}{2}\mathcal{W}^{2}\right) \epsilon ^{ac}\epsilon
^{bd}\epsilon _{a\left( c\right| }\epsilon _{b\left| d\right) }  \notag \\
&=&\mathcal{X}+\frac{1}{2}\mathcal{W}^{2}=2\left( \mathbf{I}^{\prime }-%
\mathbf{I}^{\prime \prime }\right) ,  \label{31}
\end{eqnarray}
where $\mathbf{I}^{\prime }$ and $\mathbf{I}^{\prime \prime }$ are
duality-invariant order-$4$ polynomials respectively defined by (3.12) and
(3.13) of \cite{FMOSY-1}, which we report here in current notation (recall
Footnote 3):
\begin{equation}
\left.
\begin{array}{l}
\mathbf{I}^{\prime }\equiv -\frac{1}{2}\text{Tr}_{\eta }\left(
T_{11}^{\prime }T_{22}^{\prime }\right) ; \\
\\
\mathbf{I}^{\prime \prime }\equiv -\frac{1}{2}\text{Tr}_{\eta }\left(
T_{12}^{\prime 2}\right) ;
\end{array}
\right\} \Rightarrow \mathbf{I}^{\prime }-\mathbf{I}^{\prime \prime }=\frac{1%
}{2}\left[ \text{Tr}_{\eta }\left( T_{12}^{\prime 2}\right) -\text{Tr}_{\eta
}\left( T_{11}^{\prime }T_{22}^{\prime }\right) \right] .  \label{I'-I''}
\end{equation}
By virtue of (\ref{15}) and (\ref{Chi-def}), it also holds that (\textit{cfr.%
} Eq. (4.13) of \cite{FMOSY-1}):
\begin{equation}
\mathbf{I}^{\prime }-\mathbf{I}^{\prime \prime }=\frac{1}{2}\left( \mathcal{X%
}+\frac{1}{2}\mathcal{W}^{2}\right) .
\end{equation}
Furthermore, one can derive a simple identity relating $I_{\left( ab\right)
\left( cd\right) }$ (\ref{26}) and $\mathcal{J}_{\left( ab\right) \left(
cd\right) }$ (\ref{22-2}):
\begin{equation}
\mathcal{J}_{\left( ab\right) \left( cd\right) }=I_{\left( ab\right) \left(
cd\right) }-\mathcal{W}^{2}\epsilon _{a\left( c\right| }\epsilon _{b\left|
d\right) },  \label{32}
\end{equation}
in turn implying
\begin{equation}
\mathcal{J}_{\left( abcd\right) }=I_{\left( abcd\right) }.  \label{33}
\end{equation}

\item
\begin{equation}
\left( \mathbf{1},\mathbf{3},\mathbf{Adj}\right) \times \left( \mathbf{1},%
\mathbf{3},\mathbf{Adj}\right) =\underset{\left[ SL_{h}\left( 2,\mathbb{R}%
\right) \times G_{4}\right] \text{-inv.}}{\left( \mathbf{1},\mathbf{1},%
\mathbf{1}\right) }+...  \label{34}
\end{equation}
There are no other sources of duality- (nor ``horizontal''-)invariant
polynomials involving tensor products of $\left( \mathbf{1},\mathbf{3},%
\mathbf{Adj}\right) $. The order-$4$ ``horizontal'' invariant polynomial
from (\ref{34}) can irreducibly be written as
\begin{equation}
T_{\left[ ab\right] ~\left( \alpha \beta \right) }^{\left[ \Lambda \Sigma %
\right] }T_{\left[ cd\right] ~\left( \beta \gamma \right) }^{\left[ \Xi
\Delta \right] }\eta _{\Lambda \Xi }\eta _{\Sigma \Delta }\epsilon ^{\alpha
\gamma }\epsilon ^{\beta \delta }.  \label{35}
\end{equation}
However, without any loss of generality, one can instead consider (half of)
the tensor product of the corresponding $SL_{h}\left( 2,\mathbb{R}\right) $ $%
\epsilon $-traces (which is also the unique independent manifestly $%
SL_{h}\left( 2,\mathbb{R}\right) $-invariant combination), obtaining
\begin{equation}
\frac{1}{2}T_{\left( \alpha \beta \right) }^{\left[ \Lambda \Sigma \right]
}T_{\left( \beta \gamma \right) }^{\left[ \Xi \Delta \right] }\eta _{\Lambda
\Xi }\eta _{\Sigma \Delta }\epsilon ^{\alpha \gamma }\epsilon ^{\beta \delta
}=\frac{1}{2}T_{\left[ ab\right] ~\left( \alpha \beta \right) }^{\left[
\Lambda \Sigma \right] }T_{\left[ cd\right] ~\left( \beta \gamma \right) }^{%
\left[ \Xi \Delta \right] }\eta _{\Lambda \Xi }\eta _{\Sigma \Delta
}\epsilon ^{ab}\epsilon ^{cd}\epsilon ^{\alpha \gamma }\epsilon ^{\beta
\delta }=\mathcal{X}+\frac{1}{2}\mathcal{W}^{2},  \label{36}
\end{equation}
consistently matching the result (\ref{31}), because Eqs. (\ref{31}) and (%
\ref{36}) actually share the same left-hand side.

\item
\begin{equation}
\left( \mathbf{1},\mathbf{1},\mathbf{S}_{0}\right) \times \left( \mathbf{1},%
\mathbf{1},\mathbf{S}_{0}\right) =\underset{\left[ SL_{h}\left( 2,\mathbb{R}%
\right) \times G_{4}\right] \text{-inv.}}{\left( \mathbf{1},\mathbf{1},%
\mathbf{1}\right) }+...  \label{37}
\end{equation}
Since $\mathbf{1}\notin \mathbf{S}_{2}\times \mathbf{Adj}$, there are no
other sources of duality- (nor ``horizontal''-)invariant polynomials
involving tensor products of $\left( \mathbf{1},\mathbf{1},\mathbf{S}%
_{0}\right) $. The order-$4$ ``horizontal'' invariant polynomial from (\ref
{37}) can irreducibly be written as
\begin{eqnarray}
T_{\left[ ab\right] ~\left[ \alpha \beta \right] }^{0~\left( \Lambda \Sigma
\right) }T_{\left[ ab\right] ~\left[ \alpha \beta \right] }^{0~\left( \Xi
\Delta \right) }\eta _{\Lambda \Xi }\eta _{\Sigma \Delta } &=&\left[ T_{%
\left[ ab\right] ~\left[ \alpha \beta \right] }^{\left( \Lambda \Sigma
\right) }-\frac{1}{n}\eta ^{\Lambda \Sigma }\text{Tr}_{\eta }\left( T_{\left[
ab\right] ~\left[ \alpha \beta \right] }^{\prime \prime }\right) \right]
\cdot  \notag \\
&&\cdot \left[ T_{\left[ cd\right] ~\left[ \gamma \delta \right] }^{\left(
\Xi \Delta \right) }-\frac{1}{n}\eta ^{\Lambda \Sigma }\text{Tr}_{\eta
}\left( T_{\left[ cd\right] ~\left[ \gamma \delta \right] }^{\prime \prime
}\right) \right] \eta _{\Lambda \Xi }\eta _{\Sigma \Delta },  \label{38}
\end{eqnarray}
where the $SO(2,n-2)$-matrix notation (\ref{mn-T''}) has been recalled.
However, without any loss of generality, one can instead consider the tensor
product of the corresponding $SL_{h}\left( 2,\mathbb{R}\right) $ and $%
SL_{v}\left( 2,\mathbb{R}\right) $ $\epsilon $-traces (which is also the
unique independent manifestly $SO_{h}^{v}\left( 2,2\right) $-invariant
combination), obtaining
\begin{equation}
\begin{array}{l}
T^{0~\left( \Lambda \Sigma \right) }T^{0~\left( \Xi \Delta \right) }\eta
_{\Lambda \Xi }\eta _{\Sigma \Delta }\equiv \epsilon ^{ab}\epsilon ^{\alpha
\beta }\epsilon ^{cd}\epsilon ^{\gamma \delta }T_{\left[ ab\right] ~\left[
\alpha \beta \right] }^{0~\left( \Lambda \Sigma \right) }T_{\left[ cd\right]
~\left[ \gamma \delta \right] }^{0~\left( \Xi \Delta \right) }\eta _{\Lambda
\Xi }\eta _{\Sigma \Delta } \\
~ \\
=\epsilon ^{ab}\epsilon ^{\alpha \beta }\epsilon ^{cd}\epsilon ^{\gamma
\delta }\left[ T_{\left[ ab\right] ~\left[ \alpha \beta \right] }^{\left(
\Lambda \Sigma \right) }-\frac{1}{n}\eta ^{\Lambda \Sigma }\text{Tr}_{\eta
}\left( T_{\left[ ab\right] ~\left[ \alpha \beta \right] }^{\prime \prime
}\right) \right] \cdot \\
\cdot \left[ T_{\left[ cd\right] ~\left[ \gamma \delta \right] }^{\left( \Xi
\Delta \right) }-\frac{1}{n}\eta ^{\Xi \Delta }\text{Tr}_{\eta }\left( T_{%
\left[ cd\right] ~\left[ \gamma \delta \right] }^{\prime \prime }\right)
\right] \eta _{\Lambda \Xi }\eta _{\Sigma \Delta } \\
~ \\
=\epsilon ^{ab}\epsilon ^{\alpha \beta }\epsilon ^{cd}\epsilon ^{\gamma
\delta }T_{\left[ ab\right] ~\left[ \alpha \beta \right] }^{\left( \Lambda
\Sigma \right) }T_{\left[ cd\right] ~\left[ \gamma \delta \right] }^{\left(
\Xi \Delta \right) }\eta _{\Lambda \Xi }\eta _{\Sigma \Delta } \\
-\frac{1}{n}\epsilon ^{ab}\epsilon ^{\alpha \beta }\epsilon ^{cd}\epsilon
^{\gamma \delta }\text{Tr}_{\eta }\left( T_{\left[ ab\right] ~\left[ \alpha
\beta \right] }^{\prime \prime }\right) \text{Tr}_{\eta }\left( T_{\left[ cd%
\right] ~\left[ \gamma \delta \right] }^{\prime \prime }\right) \\
~ \\
\equiv T^{\left( \Lambda \Sigma \right) }T^{\left( \Xi \Delta \right) }\eta
_{\Lambda \Xi }\eta _{\Sigma \Delta }-\frac{1}{n}\text{Tr}_{\eta }\left(
T^{\prime \prime }\right) \text{Tr}_{\eta }\left( T^{\prime \prime }\right) .
\end{array}
\label{39}
\end{equation}
Observing that the definition (\ref{15}) can be rewritten as
\begin{equation}
\text{Tr}_{\eta }\left( T^{\prime \prime }\right) =2\mathcal{W},  \label{40}
\end{equation}
definitions (\ref{T-Tensors}) imply that
\begin{equation}
T^{0~\left( \Lambda \Sigma \right) }\equiv T^{\left( \Lambda \Sigma \right)
}-\frac{2}{n}\eta ^{\Lambda \Sigma }\mathcal{W}.  \label{41}
\end{equation}
On the other hand, an explicit computation yields
\begin{equation}
T^{\left( \Lambda \Sigma \right) }T^{\left( \Xi \Delta \right) }\eta
_{\Lambda \Xi }\eta _{\Sigma \Delta }=-2\left( 2\mathcal{X}-\mathcal{W}%
^{2}\right) .  \label{42}
\end{equation}
Therefore, by inserting (\ref{40})-(\ref{42}) into (\ref{39}), the following
expression of the corresponding order-$4$ ``horizontal'' invariant
polynomial is achieved:
\begin{equation}
T^{0~\left( \Lambda \Sigma \right) }T^{0~\left( \Xi \Delta \right) }\eta
_{\Lambda \Xi }\eta _{\Sigma \Delta }=-4\mathcal{X}+\left( 2-\frac{4}{n}%
\right) \mathcal{W}^{2}.  \label{43}
\end{equation}
Note that, from (\ref{G4-red}) and observations below, the coefficient of $%
\mathcal{W}^{2}$ in (\ref{43}) is strictly positive in all CV models.

\item
\begin{equation}
\left( \mathbf{1},\mathbf{1},\mathbf{1}\right) \times \left( \mathbf{1},%
\mathbf{1},\mathbf{1}\right) =\underset{\left[ SL_{h}\left( 2,\mathbb{R}%
\right) \times G_{4}\right] \text{-inv.}}{\left( \mathbf{1},\mathbf{1},%
\mathbf{1}\right) }  \label{44}
\end{equation}
By recalling (\ref{T-Tensors}), the order-$4$ ``horizontal'' invariant
polynomial from (\ref{44}) can irreducibly be written as
\begin{equation}
\mathbf{T}_{\left[ ab\right] ~\left[ \alpha \beta \right] }^{\left( \Lambda
\Sigma \right) }\mathbf{T}_{\left[ cd\right] ~\left[ \gamma \delta \right]
}^{\left( \Xi \Delta \right) }\eta _{\Lambda \Xi }\eta _{\Sigma \Delta
}\equiv \frac{1}{n}\text{Tr}_{\eta }\left( T_{\left[ ab\right] ~\left[
\alpha \beta \right] }^{\prime \prime }\right) \text{Tr}_{\eta }\left( T_{%
\left[ cd\right] ~\left[ \gamma \delta \right] }^{\prime \prime }\right) .
\label{45}
\end{equation}
However, without any loss of generality, one can instead consider the tensor
product of the corresponding $SL_{h}\left( 2,\mathbb{R}\right) $ and $%
SL_{v}\left( 2,\mathbb{R}\right) $ $\epsilon $-traces (once again, the
unique independent manifestly $SO_{h}^{v}\left( 2,2\right) $-invariant
combination), obtaining
\begin{eqnarray}
\mathbf{T}^{\left( \Lambda \Sigma \right) }\mathbf{T}^{\left( \Xi \Delta
\right) }\eta _{\Lambda \Xi }\eta _{\Sigma \Delta } &\equiv &\epsilon
^{ab}\epsilon ^{\alpha \beta }\epsilon ^{cd}\epsilon ^{\gamma \delta }%
\mathbf{T}_{\left[ ab\right] ~\left[ \alpha \beta \right] }^{\left( \Lambda
\Sigma \right) }\mathbf{T}_{\left[ cd\right] ~\left[ \gamma \delta \right]
}^{\left( \Xi \Delta \right) }\eta _{\Lambda \Xi }\eta _{\Sigma \Delta }
\notag \\
&=&\frac{1}{n}\epsilon ^{ab}\epsilon ^{\alpha \beta }\epsilon ^{cd}\epsilon
^{\gamma \delta }\text{Tr}_{\eta }\left( T_{\left[ ab\right] ~\left[ \alpha
\beta \right] }^{\prime \prime }\right) \text{Tr}_{\eta }\left( T_{\left[ cd%
\right] ~\left[ \gamma \delta \right] }^{\prime \prime }\right)  \notag \\
&=&\frac{1}{n}\text{Tr}_{\eta }\left( T^{\prime \prime }\right) \text{Tr}%
_{\eta }\left( T^{\prime \prime }\right) =\frac{4}{n}\mathcal{W}^{2},
\label{46}
\end{eqnarray}
where Eq. (\ref{40}) was used.
\end{enumerate}

\subsubsection{\label{Order-4-Summary}Summary}

The above analysis completes, at order $4$ in the fluxes, the treatment
given in \cite{FMOSY-1} and \cite{Small-1}.

Besides $\mathcal{W}^{2}$ and $\mathcal{X}$, no other ``horizontal''
invariant homogeneous polynomials of order $4$ in the BH fluxes $\mathcal{Q}%
_{1\alpha }^{\Lambda }$ and $\mathcal{Q}_{2\alpha }^{\Lambda }$ can be
introduced.

Concerning duality-invariant homogeneous polynomials of order $4$, the
Dixmier tensor $I_{\left( abcd\right) }$ \cite{Dixmier}, sitting in the spin
$s=2$ irrep. $\mathbf{5}$ of the ``horizontal'' symmetry $SL_{h}\left( 2,%
\mathbb{R}\right) $, generally defined by (\ref{Dixmier-tensor-def}) and
present in the analysis of \cite{FMOSY-1,ADFMT-1}, is (due to (\ref{33}))
the unique algebraically independent duality-invariant tensor sitting in an
irrep. of $SL_{h}\left( 2,\mathbb{R}\right) $ itself. Other
duality-invariant tensors of mixed ``horizontal'' symmetry, such as $%
I_{\left( ab\right) \left( cd\right) }$ (\ref{26}) and $\mathcal{J}_{\left(
ab\right) \left( cd\right) }$ (\ref{22-2}) (related by (\ref{32})) can be
introduced, but they do not sit in ``horizontal'' irreps..

\subsection{\label{Order-6}Order $6$}

Since there is no duality- nor ``horizontal''- invariant polynomial
structure at order $5$, we proceed to analyze the order $6$ in the fluxes of
the $2$-centered BH solution in the framework under consideration. By
exploiting the associativity of the (ir)reps.' tensor product, in each of
the $SL_{h}\left( 2,\mathbb{R}\right) $- and $SL_{v}\left( 2,\mathbb{R}%
\right) $- sectors one gets
\begin{equation}
\mathbf{2}\times \mathbf{2}\times \mathbf{2}\times \mathbf{2}\times \mathbf{2%
}\times \mathbf{2}=\left( \mathbf{1+3+5+}2\cdot \mathbf{3}+\mathbf{1}\right)
\times \left( \mathbf{3}+\mathbf{1}\right) =5\cdot \mathbf{1}+9\cdot \mathbf{%
3}+5\cdot \mathbf{5}+\mathbf{7}.
\end{equation}
On the other hand, in the $SO\left( 2,n-2\right) $ sector the sources of
singlets list as follows:
\begin{equation}
\begin{array}{l}
\mathbf{Adj}\times \mathbf{Adj}\times \mathbf{Adj}=\mathbf{1}+...; \\
\mathbf{Adj}\times \mathbf{Adj}\times \mathbf{S}_{0}=\mathbf{1}+...; \\
\mathbf{Adj}\times \mathbf{Adj}\times \mathbf{1}=\mathbf{1}+...; \\
\mathbf{S}_{0}\times \mathbf{S}_{0}\times \mathbf{Adj}=\mathbf{1}+...; \\
\mathbf{S}_{0}\times \mathbf{S}_{0}\times \mathbf{S}_{0}=\mathbf{1}+...; \\
\mathbf{S}_{0}\times \mathbf{S}_{0}\times \mathbf{1}=\mathbf{1}+...; \\
\mathbf{1}\times \mathbf{1}\times \mathbf{1}=\mathbf{1}.
\end{array}
\end{equation}

Thus, the ``horizontal'' invariant homogeneous polynomials at order $6$ in
the fluxes arise as singlets $\left( \mathbf{1},\mathbf{1},\mathbf{1}\right)
$ among other representations from the following tensor products (in
determining the corresponding tensor structure, we will disregard the
irreducible splitting (\ref{4}), irrelevant for our purposes):

\begin{enumerate}
\item
\begin{equation}
\begin{array}{l}
\left( \mathbf{3},\mathbf{3},\mathbf{S}_{0}\right) \times \left( \mathbf{3},%
\mathbf{3},\mathbf{S}_{0}\right) \times \left( \mathbf{3},\mathbf{3},\mathbf{%
S}_{0}\right) ; \\
\left( \mathbf{3},\mathbf{3},\mathbf{S}_{0}\right) \times \left( \mathbf{3},%
\mathbf{3},\mathbf{S}_{0}\right) \times \left( \mathbf{3},\mathbf{3},\mathbf{%
1}\right) ; \\
\left( \mathbf{3},\mathbf{3},\mathbf{1}\right) \times \left( \mathbf{3},%
\mathbf{3},\mathbf{1}\right) \times \left( \mathbf{3},\mathbf{3},\mathbf{1}%
\right) ,
\end{array}
\end{equation}
whose singlets correspond to the following\footnote{%
As in all cases, the reported index-contraction structure is the unique
independent one (possibly taking into account the splitting (\ref{4}), as
well).} ``horizontal'' invariant homogeneous polynomial of order $6$:
\begin{equation}
T_{\left( ab\right) ~\left( \alpha \beta \right) }^{\left( \Lambda \Sigma
\right) }T_{\left( cd\right) ~\left( \gamma \delta \right) }^{\left( \Xi
\Delta \right) }T_{\left( ef\right) ~\left( \eta \lambda \right) }^{\left(
\Gamma \Pi \right) }\epsilon ^{af}\epsilon ^{bc}\epsilon ^{de}\epsilon
^{\alpha \lambda }\epsilon ^{\beta \gamma }\epsilon ^{\delta \eta }\eta
_{\Lambda \Pi }\eta _{\Sigma \Xi }\eta _{\Delta \Gamma }=24\mathbf{I}_{6}-24%
\mathcal{WX}+12\mathcal{W}^{3},  \label{6-1}
\end{equation}
where $\mathbf{I}_{6}$ is the order-$6$ ``horizontal'' invariant polynomial
defined by (3.16) of \cite{FMOSY-1}, which we report here in current
notation (recall Footnote 3 and notation (\ref{mn-T'})):
\begin{equation}
\mathbf{I}_{6}\equiv -\text{Tr}_{\eta }\left( T_{11}^{\prime }T_{22}^{\prime
}T_{12}^{\prime }\right) .  \label{I6}
\end{equation}

\item
\begin{equation}
\begin{array}{l}
\left( \mathbf{3},\mathbf{3},\mathbf{S}_{0}\right) \times \left( \mathbf{3},%
\mathbf{3},\mathbf{S}_{0}\right) \times \left( \mathbf{1},\mathbf{1},\mathbf{%
S}_{0}\right) ; \\
\left( \mathbf{3},\mathbf{3},\mathbf{1}\right) \times \left( \mathbf{3},%
\mathbf{3},\mathbf{1}\right) \times \left( \mathbf{1},\mathbf{1},\mathbf{1}%
\right) ; \\
\left( \mathbf{3},\mathbf{3},\mathbf{S}_{0}\right) \times \left( \mathbf{3},%
\mathbf{3},\mathbf{1}\right) \times \left( \mathbf{1},\mathbf{1},\mathbf{S}%
_{0}\right) ; \\
\left( \mathbf{3},\mathbf{3},\mathbf{S}_{0}\right) \times \left( \mathbf{3},%
\mathbf{3},\mathbf{S}_{0}\right) \times \left( \mathbf{1},\mathbf{1},\mathbf{%
1}\right) ,
\end{array}
\end{equation}
whose singlets correspond to the following ``horizontal'' invariant:
\begin{equation}
T_{\left( ab\right) ~\left( \alpha \beta \right) }^{\left( \Lambda \Sigma
\right) }T_{\left( cd\right) ~\left( \gamma \delta \right) }^{\left( \Xi
\Delta \right) }T^{\left( \Gamma \Pi \right) }\epsilon ^{ac}\epsilon
^{bd}\epsilon ^{\alpha \gamma }\epsilon ^{\beta \delta }\eta _{\Lambda \Xi
}\eta _{\Sigma \Gamma }\eta _{\Delta \Pi }=3\mathbf{I}_{6}-7\mathcal{WX}+%
\frac{5}{2}\mathcal{W}^{3}.  \label{6-2}
\end{equation}

\item
\begin{equation}
\begin{array}{l}
\left( \mathbf{3},\mathbf{3},\mathbf{S}_{0}\right) \times \left( \mathbf{3},%
\mathbf{1},\mathbf{Adj}\right) \times \left( \mathbf{1},\mathbf{3},\mathbf{%
Adj}\right) ; \\
\left( \mathbf{3},\mathbf{3},\mathbf{1}\right) \times \left( \mathbf{3},%
\mathbf{1},\mathbf{Adj}\right) \times \left( \mathbf{1},\mathbf{3},\mathbf{%
Adj}\right) ,
\end{array}
\end{equation}
whose singlets correspond to the following ``horizontal'' invariant:
\begin{equation}
T_{\left( ab\right) ~\left( \alpha \beta \right) }^{\left( \Lambda \Sigma
\right) }T_{\left( cd\right) }^{\left[ \Xi \Delta \right] }T_{\left( \eta
\lambda \right) }^{\left[ \Gamma \Pi \right] }\epsilon ^{ac}\epsilon
^{bd}\epsilon ^{\alpha \eta }\epsilon ^{\beta \lambda }\eta _{\Lambda \Pi
}\eta _{\Sigma \Xi }\eta _{\Delta \Gamma }=3\mathbf{I}_{6}+\mathcal{WX}+%
\frac{1}{2}\mathcal{W}^{3}.  \label{6-3}
\end{equation}

\item
\begin{equation}
\left( \mathbf{3},\mathbf{1},\mathbf{Adj}\right) \times \left( \mathbf{3},%
\mathbf{1},\mathbf{Adj}\right) \times \left( \mathbf{3},\mathbf{1},\mathbf{%
Adj}\right) ,
\end{equation}
whose singlet corresponds to the following ``horizontal'' invariant (recall
definition (\ref{I6})):
\begin{equation}
T_{\left( ab\right) }^{\left[ \Lambda \Sigma \right] }T_{\left( cd\right) }^{%
\left[ \Xi \Delta \right] }T_{\left( ef\right) }^{\left[ \Gamma \Pi \right]
}\epsilon ^{af}\epsilon ^{bc}\epsilon ^{de}\eta _{\Lambda \Pi }\eta _{\Sigma
\Xi }\eta _{\Delta \Gamma }=6\mathbf{I}_{6}.  \label{6-4}
\end{equation}

\item
\begin{equation}
\begin{array}{l}
\left( \mathbf{3},\mathbf{1},\mathbf{Adj}\right) \times \left( \mathbf{3},%
\mathbf{1},\mathbf{Adj}\right) \times \left( \mathbf{1},\mathbf{1},\mathbf{S}%
_{0}\right) ; \\
\left( \mathbf{3},\mathbf{1},\mathbf{Adj}\right) \times \left( \mathbf{3},%
\mathbf{1},\mathbf{Adj}\right) \times \left( \mathbf{1},\mathbf{1},\mathbf{1}%
\right) ,
\end{array}
\end{equation}
whose singlets correspond to the following ``horizontal'' invariant:
\begin{equation}
T_{\left( ab\right) }^{\left[ \Lambda \Sigma \right] }T_{\left( cd\right) }^{%
\left[ \Xi \Delta \right] }T^{\left( \Gamma \Pi \right) }\epsilon
^{ac}\epsilon ^{bd}\epsilon ^{\alpha \eta }\epsilon ^{\beta \lambda }\eta
_{\Lambda \Pi }\eta _{\Sigma \Xi }\eta _{\Delta \Gamma }=6\mathbf{I}_{6}-2%
\mathcal{WX}-\mathcal{W}^{3}.  \label{6-5}
\end{equation}

\item
\begin{equation}
\left( \mathbf{1},\mathbf{3},\mathbf{Adj}\right) \times \left( \mathbf{1},%
\mathbf{3},\mathbf{Adj}\right) \times \left( \mathbf{1},\mathbf{3},\mathbf{%
Adj}\right) ,
\end{equation}
whose singlet corresponds to the ``horizontal'' invariant (\ref{6-4}):
\begin{equation}
T_{\left( \alpha \beta \right) }^{\left[ \Lambda \Sigma \right] }T_{\left(
\gamma \delta \right) }^{\left[ \Xi \Delta \right] }T_{\left( \eta \lambda
\right) }^{\left[ \Gamma \Pi \right] }\epsilon ^{\alpha \lambda }\epsilon
^{\beta \gamma }\epsilon ^{\delta \eta }\eta _{\Lambda \Pi }\eta _{\Sigma
\Xi }\eta _{\Delta \Gamma }=6\mathbf{I}_{6},  \label{6-6}
\end{equation}
which makes the ``horizontal'' invariance of definition (\ref{I6}) manifest.

\item
\begin{equation}
\begin{array}{l}
\left( \mathbf{1},\mathbf{3},\mathbf{Adj}\right) \times \left( \mathbf{1},%
\mathbf{3},\mathbf{Adj}\right) \times \left( \mathbf{1},\mathbf{1},\mathbf{S}%
_{0}\right) ; \\
\left( \mathbf{1},\mathbf{3},\mathbf{Adj}\right) \times \left( \mathbf{1},%
\mathbf{3},\mathbf{Adj}\right) \times \left( \mathbf{1},\mathbf{1},\mathbf{1}%
\right) ,
\end{array}
\end{equation}
whose singlets correspond to the ``horizontal'' invariant (\ref{6-5}):
\begin{equation}
T_{\left( \alpha \beta \right) }^{\left[ \Lambda \Sigma \right] }T_{\left(
\gamma \delta \right) }^{\left[ \Xi \Delta \right] }T^{\left( \Gamma \Pi
\right) }\epsilon ^{\alpha \gamma }\epsilon ^{\beta \delta }\eta _{\Lambda
\Pi }\eta _{\Sigma \Xi }\eta _{\Delta \Gamma }=6\mathbf{I}_{6}-2\mathcal{WX}-%
\mathcal{W}^{3}.  \label{6-7}
\end{equation}

\item
\begin{equation}
\begin{array}{l}
\left( \mathbf{1},\mathbf{1},\mathbf{S}_{0}\right) \times \left( \mathbf{1},%
\mathbf{1},\mathbf{S}_{0}\right) \times \left( \mathbf{1},\mathbf{1},\mathbf{%
S}_{0}\right) ; \\
\left( \mathbf{1},\mathbf{1},\mathbf{S}_{0}\right) \times \left( \mathbf{1},%
\mathbf{1},\mathbf{S}_{0}\right) \times \left( \mathbf{1},\mathbf{1},\mathbf{%
1}\right) ; \\
\left( \mathbf{1},\mathbf{1},\mathbf{1}\right) \times \left( \mathbf{1},%
\mathbf{1},\mathbf{1}\right) \times \left( \mathbf{1},\mathbf{1},\mathbf{1}%
\right) ,
\end{array}
\end{equation}
whose singlets correspond to the following ``horizontal'' invariant:
\begin{equation}
T^{\left( \Lambda \Sigma \right) }T^{\left( \Xi \Delta \right) }T^{\left(
\Gamma \Pi \right) }\eta _{\Lambda \Pi }\eta _{\Sigma \Xi }\eta _{\Delta
\Gamma }=12\mathbf{I}_{6}-12\mathcal{WX}+2\mathcal{W}^{3}.  \label{6-8}
\end{equation}

\item
\begin{equation}
\left( \mathbf{3},\mathbf{3},\mathbf{S}_{0}\right) \times \left( \mathbf{3},%
\mathbf{3},\mathbf{S}_{0}\right) \times \left( \mathbf{3},\mathbf{1},\mathbf{%
Adj}\right) ,
\end{equation}
whose singlet corresponds to the following ``horizontal'' invariant:
\begin{equation}
T_{\left( ab\right) ~\left( \alpha \beta \right) }^{\left( \Lambda \Sigma
\right) }T_{\left( cd\right) ~\left( \gamma \delta \right) }^{\left( \Xi
\Delta \right) }T_{\left( ef\right) }^{\left[ \Gamma \Pi \right] }\epsilon
^{\alpha f}\epsilon ^{bc}\epsilon ^{de}\epsilon ^{\alpha \gamma }\epsilon
^{\beta \delta }\eta _{\Lambda \Pi }\eta _{\Sigma \Xi }\eta _{\Delta \Gamma
}=3\mathbf{I}_{6}-2\mathcal{WX}-\mathcal{W}^{3}.  \label{6-9}
\end{equation}

\item
\begin{equation}
\left( \mathbf{3},\mathbf{3},\mathbf{S}_{0}\right) \times \left( \mathbf{3},%
\mathbf{3},\mathbf{S}_{0}\right) \times \left( \mathbf{1},\mathbf{3},\mathbf{%
Adj}\right) ,
\end{equation}
whose singlet corresponds to the ``horizontal'' invariant (\ref{6-9}):
\begin{equation}
T_{\left( ab\right) ~\left( \alpha \beta \right) }^{\left( \Lambda \Sigma
\right) }T_{\left( cd\right) ~\left( \gamma \delta \right) }^{\left( \Xi
\Delta \right) }T_{\left( \eta \lambda \right) }^{\left[ \Gamma \Pi \right]
}\epsilon ^{\alpha c}\epsilon ^{bd}\epsilon ^{\alpha \lambda }\epsilon
^{\beta \gamma }\epsilon ^{\delta \eta }\eta _{\Lambda \Pi }\eta _{\Sigma
\Xi }\eta _{\Delta \Gamma }=3\mathbf{I}_{6}-2\mathcal{WX}-\mathcal{W}^{3}.
\label{6-10}
\end{equation}
\medskip \smallskip
\end{enumerate}

\subsubsection{\label{Order-6-Summary}Summary}

The above analysis completes, at order $6$ in the fluxes, the treatment
given in \cite{FMOSY-1} and \cite{Small-1}.

Besides $\mathcal{W}^{3}$ and $\mathcal{WX}$, also the ``horizontal''
invariant $\mathbf{I}_{6}$ (\ref{I6}) can be introduced. Concerning this, it
is worth recalling here that two ``horizontal'' invariant order-$6$
homogeneous polynomials can be naturally introduced in CV models:

\begin{itemize}
\item  the $\mathbf{I}_{6}$ defined by (\ref{I6}) above \cite{FMOSY-1};

\item  the $\mathbf{I}_{6}^{\prime }$ given by (3.11) and (3.24) of \cite
{ADFMT-1}, whose manifestly ``horizontal''-invariant formulation in the CV
symplectic frame \cite{CDFVP-1} reads
\begin{eqnarray}
\mathbf{I}_{6}^{\prime } &\equiv &\frac{1}{2}\mathbb{C}_{\alpha \beta
}^{\Lambda \Sigma }\widetilde{\mathcal{Q}}_{\Lambda \mid abc}^{\alpha }%
\widetilde{\mathcal{Q}}_{\Sigma \mid def}^{\beta }\epsilon ^{ad}\epsilon
^{be}\epsilon ^{cf},  \label{spin=2-CV} \\
\widetilde{\mathcal{Q}}_{\Lambda \mid abc}^{\alpha } &\equiv &\frac{1}{2}%
\mathbb{K}_{\Lambda \Sigma \Xi \Delta }^{\alpha \beta \gamma \delta }%
\mathcal{Q}_{a\beta }^{\Sigma }\mathcal{Q}_{b\gamma }^{\Xi }\mathcal{Q}%
_{c\delta }^{\Delta },  \label{spin=3/2-CV}
\end{eqnarray}
where, the symplectic-invariant $\mathbb{C}$-structure reads
\begin{equation}
\mathbb{C}^{MN}=\mathbb{C}_{\alpha \beta }^{\Lambda \Sigma }=\eta ^{\Lambda
\Sigma }\epsilon _{\alpha \beta },  \label{C-structure}
\end{equation}
consistent with the CV splitting (\ref{CV-split}).
\end{itemize}

The general\footnote{%
As discussed in \cite{ADFMT-1} and in \cite{Small-1}, an important
difference between CV models and those $D=4$ symmetric models with simple $U$%
-duality groups (named \textit{irreducible} symmetric models in such Refs.)
is that in these latter $\mathcal{X}$ vanishes identically (due to the
holding of Eq. (3.7) of \cite{ADFMT-1}).} relation between $\mathbf{I}_{6}$
and $\mathbf{I}_{6}^{\prime }$ is discussed in Sec. 3 of \cite{Small-1}; in
CV models, such a relation is given by (also recall (\ref{31}) and (\ref
{I'-I''}))
\begin{equation}
\mathbf{I}_{6}=\mathbf{I}_{6}^{\prime }-\frac{1}{6}\mathcal{X}\mathcal{W}-%
\frac{1}{12}\mathcal{W}^{3}=\mathbf{I}_{6}^{\prime }-\frac{1}{3}\left\|
\mathbb{T}\right\| ^{2}\mathcal{W},
\end{equation}
which, besides (\ref{6-6}), provides a manifestly ``horizontal''-invariant
expression of $\mathbf{I}_{6}$ in the CV symplectic frame.

\section{\label{Hor-U(n,m)}Invariants of Pseudo-Unitary $U$-Duality}

We now discuss the ``horizontal'' symmetry of $D=4$ supergravity models with
$U$-duality group $G_{4}$ given by the pseudo-unitary group $U\left(
r,s\right) $ for some $r$ and $s$. Confining ourselves to theories with
symmetric scalar manifolds, these supergravity theories are:

\begin{itemize}
\item  the $\mathcal{N}=2$ \textit{minimally coupled} Maxwell-Einstein
theory \cite{Luciani,Gnecchi-1} ($r=1$), with scalar manifold
\begin{equation}
M_{\mathcal{N}=2}=\frac{U\left( 1,s\right) }{U\left( 1\right) \times U(s)}%
\sim \frac{SU\left( 1,s\right) }{SU\left( s\right) \times U(1)}\equiv
\mathbb{CP}^{s}  \label{N=2-mc}
\end{equation}
and vector $2$-form field strengths sitting in the (complex) fundamental
irrep. $\mathbf{s+1}$ of $G_{4}=U(1,s)$;

\item  the $\mathcal{N}=3$ matter-coupled theory \cite{N=3} ($r=3$), with
scalar manifold
\begin{equation}
M_{\mathcal{N}=3}=\frac{U\left( 3,s\right) }{U\left( 1\right) \times S\left(
U\left( 3\right) \times U(s)\right) }\sim \frac{SU\left( 3,s\right) }{%
SU\left( 3\right) \times SU(s)\times U\left( 1\right) },  \label{N=3}
\end{equation}
and vector $2$-form field strengths sitting in the (complex) fundamental
irrep. $\mathbf{s+3}$ of $G_{4}=U(3,s)$.
\end{itemize}

It is here worth recalling that $\mathcal{N}=2$ supergravity \textit{%
minimally coupled} to $3$ vector multiplets is \textit{``bosonic twin''} to
\textit{(i.e.} shares the very same bosonic sector of) $\mathcal{N}=3$
supergravity coupled to $1$ vector multiplet \cite{Gnecchi-1,Duff-Ferrara-1}
(for a discussion of split flows and marginal stability in extended $D=4$
supergravities, see \textit{e.g.} \cite{MS-FM-1}).

As observed in \cite{MS-FMO-1} (in which the split attractor flow and
marginal stability features of theories (\ref{N=2-mc}) were investigated),
the presence of an ``extra'' $U(1)$ symmetry, acting only on the
complex(ified) flux vector $\mathcal{Q}$ but not on the (complex) scalar
fields (see \textit{e.g.} (2.35)-(2.36) of \cite{MS-FMO-1}) is due to the
fact that such theories are the only ones in which the \textit{``pure''}
theory limit (corresponding to the case in which only the graviton multiplet
present) can be obtained by simply setting to zero the number $s$ of matter
(Abelian vector) multiplets. As such, the ``extra'' $U(1)$ global factor%
\footnote{\textit{At least} for $p=2$, the relevance of the ``extra'' $%
U\left( 1\right) $ factor for the counting of $U\left( r,s\right) $%
-invariant polynomials has been discussed at the end of page $5$ of \cite
{MS-FMO-1}.} (which is not a global isometry of the scalar manifold) is
nothing but the $U(1)$ symmetry gauged by the complex scalars, which becomes
global in their absence \cite{FSZ} (recall that the $\mathcal{N}=2$ and $%
\mathcal{N}=3$ graviton multiplets do not contain scalar fields). Moreover,
in the $\mathcal{N}=2$\ case, such a $U(1)$ can also be interpreted as the
symmetry of the graviphotonic electro-magnetic system.

\subsection{$SL\left( p,\mathbb{C}\right) \times U(r,s)$}

Before dealing with the actual ``horizontal'' symmetry of these theories, it
is instructive to consider the group
\begin{equation}
SL\left( p,\mathbb{C}\right) \times U(r,s),
\end{equation}
and the orbits of complex vectors $\mathcal{V}_{i}^{A}$ ($A=1,...,r+s$, $%
i=1,...,p$) in the complex bi-fundamental irrep. $\left( \mathbf{p},\mathbf{%
r+s}\right) $ of $SL\left( p,\mathbb{C}\right) \times U(r,s)$. This
treatment can be considered the ``complexified version'' of the treatment
given in the second part of Sec. 4 of \cite{ADFMT-1}, and also the results
will be analogous.

There are only $p^{2}$ algebraically-independent $U(r,s)$-invariant
homogeneous polynomials, all of order $2$ in the fluxes, given by
\begin{equation}
\mathbf{U}_{i\overline{j}}\equiv \mathcal{V}_{i}^{A}\overline{\mathcal{V}}_{%
\overline{j}}^{\overline{B}}\eta _{A\overline{B}}\equiv \mathcal{V}_{i}\cdot
\overline{\mathcal{V}}_{\overline{j}},  \label{ni-i-jbar}
\end{equation}
where ``$\cdot $'' denotes the scalar product determined by the
pseudo-Euclidean metric $\eta _{A\overline{B}}$ of $U\left( r,s\right) $. By
respectively denoting with $\mathbf{I}_{p}$ and $\frak{G}_{p}$ the dimension
of a complete basis of $U(r,s)$-invariant polynomials and the orbit of the
irrep. $\mathbf{r+s}$ of $U(r,s)$, the counting
\begin{equation}
\mathbf{I}_{p}=p^{2}  \label{C-1}
\end{equation}
is consistent with the general counting rule \cite{FMOSY-1,ADFMT-1}:
\begin{equation}
\mathbf{I}_{p}=2\left( r+s\right) p-\text{dim}_{\mathbb{R}}\left( \frak{G}%
_{p}\right) ,
\end{equation}
because $\frak{G}_{p}$ generally is a suitable non-compact form of the
compact coset
\begin{equation}
\frak{G}_{p,\text{compact}}=\frac{U\left( r+s\right) }{U\left( r+s-p\right) }%
\text{,~dim}_{\mathbb{R}}=2\left( r+s\right) p-p^{2}.  \label{compact-G_p}
\end{equation}

On the other hand, out of the $p^{2}$ order-$2$ $U\left( r,s\right) $%
-invariant polynomials $\mathbf{U}_{i\overline{j}}$ (\ref{ni-i-jbar}), one
can construct a unique algebraically-independent $\left[ SL\left( p,\mathbb{C%
}\right) \times U(r,s)\right] $- invariant homogeneous polynomial, of order $%
2p$, defined as
\begin{equation}
\text{det}\widehat{\mathbf{G}}=\mathcal{V}_{i_{1}\overline{j}_{1}}\mathcal{V}%
_{i_{2}\overline{j}_{2}}...\mathcal{V}_{i_{p}\overline{j}_{p}}\epsilon
^{i_{1}i_{2}...i_{p}}\epsilon ^{\overline{j}_{1}\overline{j}_{2}...\overline{%
j}_{p}}.  \label{I_2p}
\end{equation}
The notation ``det$\widehat{\mathbf{G}}$'' indicates the fact that the
``horizontal''-invariant polynomial (\ref{I_2p}) is the determinant of the
Hermitian-analogue of the Gramian matrix $\mathbf{G}$ introduced in
(8.4)-(8.5) of \cite{FMOSY-1} (see also the treatment of Sec. 8\ therein).
In the $2$-centered case ($p=2$), (\ref{I_2p}) reduces to
\begin{equation}
\text{det}\widehat{\mathbf{G}}=\mathcal{V}_{i_{1}\overline{j}_{1}}\mathcal{V}%
_{i_{2}\overline{j}_{2}}\epsilon ^{i_{1}i_{2}}\epsilon ^{\overline{j}_{1}%
\overline{j}_{2}}=2\left( \left| \mathcal{V}_{1}\right| ^{2}\left| \mathcal{V%
}_{2}\right| ^{2}-\left| \mathcal{V}_{1}\cdot \overline{\mathcal{V}}_{%
\overline{2}}\right| ^{2}\right) ,
\end{equation}
which can be recognized as (twice the) the squared norm of the $\left(
SL\left( 2,\mathbb{C}\right) \sim SO\left( 3,1\right) \right) $-vector $%
\mathcal{V}_{i\overline{j}}$. By respectively denoting with $\frak{I}_{p}$
and $\widetilde{\frak{G}}_{p}$ the dimension of a complete basis of $\left[
SL\left( p,\mathbb{C}\right) \times U(r,s)\right] $-invariant polynomials
and the orbit of the irrep. $\left( \mathbf{p},\mathbf{r+s}\right) $ of $%
\left[ SL\left( p,\mathbb{C}\right) \times U(r,s)\right] $ itself, the
counting
\begin{equation}
\frak{I}_{p}=1
\end{equation}
is consistent with the general counting rule:
\begin{equation}
\frak{I}_{p}=2\left( r+s\right) p-\text{dim}_{\mathbb{R}}\left( \widetilde{%
\frak{G}}_{p}\right) ,
\end{equation}
because $\widetilde{\frak{G}}_{p}$ generally is the direct product of the
Riemannian symmetric non-compact coset ($SU\left( p\right) =mcs\left[
SL\left( p,\mathbb{C}\right) \right] $)
\begin{equation}
\frac{SL\left( p,\mathbb{C}\right) }{SU\left( p\right) },~\text{dim}_{%
\mathbb{R}}=p^{2}-1  \label{hor-coset}
\end{equation}
and of a suitable non-compact form of the compact coset (\ref{compact-G_p}):
\begin{equation}
\widetilde{\frak{G}}_{p,\text{compact}}=\frac{SL\left( p,\mathbb{C}\right) }{%
SU\left( p\right) }\times \frac{U\left( r+s\right) }{U\left( r+s-p\right) }%
\sim \frac{SL\left( p,\mathbb{C}\right) \times U\left( r+s\right) }{SU\left(
p\right) \times U\left( r+s-p\right) },~dim_{\mathbb{R}}=2\left( r+s\right)
p-1.
\end{equation}
\medskip

It is immediate to realize that $SL\left( p,\mathbb{C}\right) $ cannot be
the ``horizontal'' symmetry of a $p$-centered BH solution of the models
under consideration. Indeed, the total symmetry $SL\left( p,\mathbb{C}%
\right) \times U\left( r+s\right) $ exhibits no invariants of order $2$ in
charges, as instead the symplectic product $\mathcal{W}$ (\ref{15-bis})
should generally be.

\subsection{$SL_{h}\left( 2,\mathbb{R}\right) \times U(r,s)$}

The number and the structure of $p$-centered ($2\leqslant p\leqslant r+s$)
algebraically independent duality-invariant homogeneous polynomials in the $%
\mathcal{N}=2$ \textit{minimally coupled} theory have been already discussed
in Sec. 4.2.1 of \cite{MS-FMO-1}. We now give a unified treatment (holding
for both the theories (\ref{N=2-mc}) and (\ref{N=3})) of both $p$-centered ($%
2\leqslant p\leqslant r+s$) ``horizontal''- and duality- invariant
polynomials.

As mentioned above, the $2$-form field strengths and their dual (and thus
the corresponding fluxes) sit in the complex fundamental irrep. $\mathbf{r+s}
$ of $G_{4}=U\left( r,s\right) $. When considering a $p$-centered BH
solution, the complex(ified) flux vector $\mathcal{Q}_{i}^{A}$ ($A=1,...,r+s$%
, $i=1,...,p$) sits in the bi-fundamental irrep. $\left( \mathbf{p},\mathbf{%
r+s}\right) $ of the
\begin{equation}
\text{``horizontal''}\times U\text{-duality~group}:SL_{h}\left( p,\mathbb{R}%
\right) \times U(r,s).
\end{equation}
As noticed in \cite{MS-FMO-1} for the $\mathcal{N}=2$ \textit{minimally
coupled} case, there are only $p^{2}$ algebraically-independent $U(r,s)$%
-invariant homogeneous polynomials, all of order $2$ in the fluxes, given by
\begin{equation}
v_{ij}\equiv \mathcal{Q}_{i}^{A}\overline{\mathcal{Q}}_{j}^{\overline{B}%
}\eta _{A\overline{B}}\equiv \mathcal{Q}_{i}\cdot \overline{\mathcal{Q}}_{j}=%
\mathcal{S}_{ij}+\mathcal{W}_{ij},  \label{ni-i-jbar-2}
\end{equation}
where ``$\cdot $'' denotes the scalar product determined by the
pseudo-Euclidean metric $\eta _{A\overline{B}}$ of $U\left( r,s\right) $,
and
\begin{eqnarray}
\mathcal{S}_{ij} &\equiv &\mathcal{Q}_{(i}\cdot \overline{\mathcal{Q}}_{j)};
\label{S_ij} \\
\mathcal{W}_{ij} &\equiv &\mathcal{Q}_{[i}\cdot \overline{\mathcal{Q}}_{j]}.
\label{W_ij}
\end{eqnarray}
Note that, with respect to (\ref{ni-i-jbar}), the ``horizontal'' indices $i$%
's here belong to the real fundamental irrep. $\mathbf{p}$ of $SL_{h}\left(
p,\mathbb{R}\right) $. By respectively denoting with $I_{p}$ and $\frak{G}%
_{p}$ the dimension of a complete basis of duality invariant polynomials and
the orbit of the irrep. $\mathbf{r+s}$ of $G_{4}$, one obtains the very same
counting given by Eqs. (\ref{C-1})-(\ref{compact-G_p}).

Let us now consider the issue of ``horizontal'' invariants in the $2$%
-centered ($p=2$) case.

In this case, there are $p^{2}=4$ order-$2$ $U\left( r,s\right) $-invariant
polynomials $v_{i\overline{j}}$ (\ref{ni-i-jbar-2}), namely \cite{MS-FMO-1}:
\begin{equation}
\begin{array}{l}
\mathcal{S}_{11}=\mathcal{Q}_{1}\cdot \overline{\mathcal{Q}}_{1}\equiv
\left| \mathcal{Q}_{1}\right| ^{2}\equiv I_{2}\left( \mathcal{Q}_{1}\right) ;
\\
\\
\mathcal{S}_{22}=\left| \mathcal{Q}_{2}\right| ^{2}\equiv I_{2}\left(
\mathcal{Q}_{2}\right) ; \\
\\
\mathcal{S}_{12}=\text{Re}\left( \mathcal{Q}_{1}\cdot \overline{\mathcal{Q}}%
_{2}\right) \equiv \mathbf{I}_{s}; \\
\\
\mathcal{W}_{12}=i\text{Im}\left( \mathcal{Q}_{1}\cdot \overline{\mathcal{Q}}%
_{2}\right) \equiv i\mathbf{I}_{a}=-i\mathcal{W},
\end{array}
\label{p=2-Invs}
\end{equation}
where $i,j=1,2$, and $I_{2}\left( \mathcal{Q}\right) $ is the unique
algebraically-independent $1-$centered $U\left( r,s\right) $-invariant
polynomial (homogeneous of order-$2$ in the charges $\mathcal{Q}$'s; see
\textit{e.g.} \cite{ADF-U-Duality-d=4,Gnecchi-1}, and Refs. therein):
\begin{equation}
I_{2}\left( \mathcal{Q}\right) \equiv \mathcal{Q}^{A}\overline{\mathcal{Q}}^{%
\overline{B}}\eta _{A\overline{B}}\equiv \mathcal{Q}\cdot \overline{\mathcal{%
Q}}\equiv \left| \mathcal{Q}\right| ^{2}.  \label{Inv-quadr}
\end{equation}

Out of (\ref{p=2-Invs}), one can easily determine that the \textit{``minimal
degree''} basis of homogeneous\linebreak\ $\left[ SL_{h}\left( 2,\mathbb{R}%
\right) \times U\left( r+s\right) \right] $-invariant polynomials is
composed by one invariant of order $2$, namely $\mathcal{W}$ (\ref{15-bis}),
and by the following invariant of order $4$ ($\epsilon ^{12}\equiv 1$):
\begin{equation}
\mathcal{I}_{1}\equiv \mathcal{S}_{ij}\mathcal{S}_{kl}\epsilon ^{ik}\epsilon
^{jl}=2\left( \mathcal{S}_{11}\mathcal{S}_{22}-\mathcal{S}_{12}^{2}\right) =2%
\left[ I_{2}\left( \mathcal{Q}_{1}\right) I_{2}\left( \mathcal{Q}_{2}\right)
-\mathbf{I}_{s}^{2}\right] .
\end{equation}
In order to show this, we start and compute
\begin{equation}
v_{ij}v_{kl}\epsilon ^{ik}\epsilon ^{jl}=\left( \mathcal{Q}_{i}\cdot
\overline{\mathcal{Q}}_{j}\right) \left( \mathcal{Q}_{k}\cdot \overline{%
\mathcal{Q}}_{l}\right) \epsilon ^{ik}\epsilon ^{jl}=\left( \mathcal{S}_{ij}+%
\mathcal{W}_{ij}\right) \left( \mathcal{S}_{kl}+\mathcal{W}_{kl}\right)
\epsilon ^{ik}\epsilon ^{jl}=\mathcal{V}_{1}+2\mathcal{V}_{2}+\mathcal{V}%
_{3}.
\end{equation}
By using the Schouten identities for $SL_{h}\left( 2,\mathbb{R}\right) $%
\begin{equation}
\delta _{\lbrack a}^{e}\epsilon _{cd]}=0,
\end{equation}
it is immediate to obtain
\begin{eqnarray}
\mathcal{V}_{2} &\equiv &\mathcal{S}_{ij}\mathcal{W}_{kl}\epsilon
^{ik}\epsilon ^{jl}=0; \\
\mathcal{V}_{3} &\equiv &\mathcal{W}_{ij}\mathcal{W}_{kl}\epsilon
^{ik}\epsilon ^{jl}=2\mathcal{W}^{2},
\end{eqnarray}
such that
\begin{equation}
v_{ij}v_{kl}\epsilon ^{ik}\epsilon ^{jl}=2\left[ I_{2}\left( \mathcal{Q}%
_{1}\right) I_{2}\left( \mathcal{Q}_{2}\right) -\mathbf{I}_{s}^{2}\right] +2%
\mathcal{W}^{2}=2\left( \left| \mathcal{Q}_{1}\right| ^{2}\left| \mathcal{Q}%
_{2}\right| ^{2}-\left| \mathcal{Q}_{1}\cdot \overline{\mathcal{Q}}%
_{2}\right| ^{2}\right) .
\end{equation}
By respectively denoting with $\widehat{\frak{I}}_{2}$ and $\widehat{\frak{G}%
}_{2}$ the dimension of a complete basis of $\left[ SL\left( 2,\mathbb{R}%
\right) \times U(r,s)\right] $-invariant polynomials and the orbit of the
irrep. $\left( \mathbf{2},\mathbf{r+s}\right) $ of $\left[ SL\left( 2,%
\mathbb{R}\right) \times U(r,s)\right] $ itself, the counting
\begin{equation}
\widehat{\frak{I}}_{2}=2
\end{equation}
is consistent with the general counting rule:
\begin{equation}
\widehat{\frak{I}}_{2}=4\left( r+s\right) -\text{dim}_{\mathbb{R}}\left(
\widehat{\frak{G}}_{2}\right) ,
\end{equation}
because $\widehat{\frak{G}}_{2}$ generally is the direct product of the
Riemannian symmetric ``horizontal'' non-compact coset ($SO\left( 2\right)
=mcs\left[ SL\left( 2,\mathbb{R}\right) \right] $)
\begin{equation}
\frac{SL\left( 2,\mathbb{R}\right) }{SO\left( 2\right) },~\text{dim}_{%
\mathbb{R}}=2,
\end{equation}
and of a suitable non-compact form of the compact coset (\ref{compact-G_p}):
\begin{equation}
\widehat{\frak{G}}_{2,\text{compact}}=\frac{SL\left( 2,\mathbb{R}\right) }{%
SO\left( 2\right) }\times \frac{U\left( r+s\right) }{U\left( r+s-2\right) }%
\sim \frac{SL\left( 2,\mathbb{R}\right) \times U\left( r+s\right) }{SO\left(
2\right) \times U\left( r+s-2\right) },~dim_{\mathbb{R}}=4\left( r+s\right)
-2.
\end{equation}
\medskip

Thus, \textit{at least} in the $p=2$ case, an important feature of the
models under consideration is that the ``horizontal'' sector coset (\ref
{hor-coset}) has a non-trivial stabilizer $SO\left( 2\right) $, differently
\textit{e.g.} from the CV (\textit{alias} \textit{reducible} symmetric) \cite
{FMOSY-1,Small-1} and from the \textit{irreducible} symmetric \cite
{ADFMT-1,Exc-Reds,Small-1} models (also recall Footnote 4).

We leave the detailed investigation of the cases $p\geqslant 3$ for future
further study.

\section{\label{Conclusion}Remarks}

\subsection{\label{Remarks-CV}On CV Models}

We have given a complete analysis of the $\left[ SL_{v}\left( 2,\mathbb{R}%
\right) \times SO\left( m,n\right) \right] $- (\textit{i.e.} duality-)
and\linebreak\ $\left[ SL_{h}\left( 2,\mathbb{R}\right) \times SL_{v}\left(
2,\mathbb{R}\right) \times SO\left( m,n\right) \right] $- (\textit{i.e.}
``horizontal'') invariant homogeneous polynomials in Calabi-Vesentini (CV) $%
D=4$ supergravity models (cfr. Eq. (\ref{G4-red})), up to order $6$ in the
fluxes $\mathcal{Q}$'s included. This analysis refines and completes the
treatments of \cite{FMOSY-1,ADFMT-1,Small-1}.

Consistent with analysis of \cite{FMOSY-1} (and with the general results of
\cite{Kac}), a complete basis of homogeneous ``horizontal'' invariant
polynomials for the CV models is given by (\textit{cfr.} Eq. (8.2) of \cite
{FMOSY-1}, as well as the treatment of Sec. 4 of \cite{Small-1})
\begin{equation}
\left\{ \mathcal{W},~\mathcal{X},~\mathbf{I}_{6},\text{Tr}\left( \frak{I}%
_{0}^{2}\right) \right\} ,  \label{hor-basis-md}
\end{equation}
where the order-$8$ ``horizontal'' invariant polynomial Tr$\left( \frak{I}%
_{0}^{2}\right) $ is defined by Eq. (4.9) of \cite{FMOSY-1}. It is worth
remarking that, as yielded by the general analysis of \cite{Kac}, besides
being \textit{``of minimal order''} in the fluxes $\mathcal{Q}_{1\alpha
}^{\Lambda }$ and $\mathcal{Q}_{2\beta }^{\Sigma }$ of the two BH centers,
the basis (\ref{hor-basis-md}) is also \textit{freely generating} the ring
of ``horizontal'' invariant (homogeneous) polynomials: in other words, all
other ``horizontal'' invariant polynomials are themselves polynomials in $%
\mathcal{W}$, $\mathcal{X}$, $\mathbf{I}_{6}$ and Tr$\left( \frak{I}%
_{0}^{2}\right) $, with no syzygial constraints\footnote{%
It should be pointed out that, due to the order-$12$ syzygial constraint
given by Eq. (5.6) of \cite{FMOSY-1} (holding in all CV models), the
``horizontal''-invariant basis (\textit{cfr.} Eq. (8.1) of \cite{FMOSY-1},
as well as the treatment of Sec. 4 of \cite{Small-1})
\begin{equation*}
\left\{ \mathcal{W},~\mathcal{X},~\text{Tr}\left( \frak{I}_{0}^{2}\right) ,%
\text{Tr}\left( \frak{I}_{0}^{3}\right) \right\}
\end{equation*}
is not \textit{freely generating}.
\par
For \textit{irreducible} symmetric models \cite{ADFMT-1,Exc-Reds}, due to
the vanishing of $\mathcal{X}$ mentioned above, the complete basis \textit{%
``of minimal order''} in the $\mathcal{Q}$'s for ``horizontal''-invariant
(homogeneous) polynomials is given by
\begin{equation*}
\left\{ \mathcal{W},~\mathbf{I}_{6},\text{Tr}\left( \frak{I}_{0}^{2}\right)
,~\text{Tr}\left( \frak{I}_{0}^{3}\right) \right\} ,
\end{equation*}
and it is \textit{freely generating} \cite{Kac}. Therefore, apart from the
peculiar case of the so-called $t^{3}$ model (treated in Sec. 7 and App. B
of \cite{FMOSY-1}, as well as in Sec. 5.2 of \cite{Small-1}), the $\mathcal{X%
}=0$ limit of the order-$12$ constraint given by Eq. (5.6) of \cite{FMOSY-1}
does not hold in \textit{irreducible} symmetric models.}.

As for purely duality-invariant polynomials, we observe that their analysis
at order-$6$ in the fluxes has not been performed in Sec. \ref{Order-6}.
This is due to the fact that, from the treatment given in Sec. 5 of \cite
{FMOSY-1}, it is known that a(n in general \textit{non-freely generating})
complete basis \textit{``of minimal order''} in the $\mathcal{Q}$'s for the
purely duality-invariant (homogeneous) polynomials is\footnote{%
Of course, other choices are possible; see \textit{e.g.} Sec. 4 of \cite
{Small-1}.
\par
For \textit{irreducible} symmetric models, due to the vanishing of $\mathcal{%
X}$ mentioned above, the (in general \textit{non-freely generating})
complete basis \textit{``of minimal order''} in the $\mathcal{Q}$'s for the
purely duality-invariant (homogeneous) polynomials is given by Eq. (1.16) of
\cite{Small-1} (also in this case, other choice are of course possible; see
\textit{e.g.} Sec. 5 of \cite{Small-1}).} (\textit{cfr.} Eq. (1.15) of \cite
{Small-1})
\begin{equation}
\left\{ \mathcal{W},~\mathcal{X},~I_{\left( abcd\right) }\right\} ,
\end{equation}
and thus one does not need to seek for order-$6$ (and/or higher) purely
duality-invariant homogeneous polynomials.

We would like also to point out that the non-generic example of CV model
provided by the so-called $st^{2}$ model is treated in Sec. 6 of \cite
{FMOSY-1} and in Sec. 4.2 and App. B of \cite{Small-1}. Moreover, we recall
that the so-called $t^{3}$ model (treated in Sec. 7 of \cite{FMOSY-1} and in
Sec. 5.2 and App. B of \cite{Small-1}) is, as it is well known, an isolated
case in the classification of symmetric special K\"{a}hler geometries (see
\textit{e.g.} \cite{CFG}, and \cite{LA08-Proc} for a list of Refs.); as
such, it does not belong to the CV models (\ref{G4-red}). However, as shown
in app. B of \cite{FMOSY-1}, it can be reformulated in terms of a
``constrained'' CV symplectic frame.

\subsection{\label{Remarks-U(n,m)}On Pseudo-Unitary $U$-Duality}

On the other hand, the analysis carried out in Sec. \ref{Hor-U(n,m)} (in
turn refining the treatment given in Sec. 4.2.1 of \cite{MS-FMO-1}) yields
that the (symmetric) $D=4$ supergravity models with $U$-duality group $G_{4}$
given by the pseudo-unitary group $U\left( r,s\right) $ have a much simpler
case study concerning the duality- and ``horizontal''- invariant homogeneous
polynomials constructed out of the BH fluxes' irrep., \textit{at least} for $%
p=2$.

It is here worth commenting also about the $\mathcal{N}=2$, $D=4$ ``magic''
Maxwell-Einstein supergravity based on the \textit{irreducible} cubic Jordan
algebra $J_{3}^{\mathbb{C}}$ \cite{GST-1,GST-2} (see also \textit{e.g.} \cite
{Small-Orbits-Physics,Small-Orbits-Maths} for a recent account) and on the $%
\mathcal{N}=5$, $D=4$ \textit{``pure''} theory \cite{N=5}. Despite the fact
that their $U$-duality groups are suitable non-compact forms of the (\textit{%
special}) pseudo-unitary group $SU\left( 6\right) $ (namely, $SU(3,3)$
respectively $SU\left( 1,5\right) $), these theories do not belong to the
class of models treated in Sec. \ref{Hor-U(n,m)}. Indeed, they do not have
an ``extra'' global symmetry $U(1)$ under which their $2$-form field
strengths' fluxes are charged; this is also related to the fact that their
magnetic and electric fluxes sit in a \textit{self-real} irrep., namely the
rank-$3$ completely antisymmetric $\mathbf{20}$ of $SU\left( 6\right) $ (and
not in the complex fundamental irrep. $\mathbf{6}$ of the analogue would-be
model of the type treated in\ Sec. \ref{Hor-U(n,m)}).

It should also be noticed that (non-compact forms of) $\mathbb{CP}^{n}$
spaces as moduli spaces of string compactifications have appeared in the
literature, either as particular subspaces of complex structure deformations
of certain Calabi-Yau manifold \cite{CDR,Dixon:1989fj} or as moduli spaces
of some asymmetric orbifolds of Type II superstrings \cite{FK}--\nocite
{FF,DH}\cite{KK}, or of orientifolds \cite{Frey}.

Finally, we observe that the $D=4$ supergravity models considered in Sec.
\ref{Hor-U(n,m)} are not included in the analysis of \cite{Kac}. In fact,
only the ``real (pseudo-orthogonal) analogues'' of such models (in which the
analogue of $\mathcal{W}_{ij}$ vanishes; see the treatment given in the
second part of Sec. 4 of \cite{ADFMT-1}) can be found in Table II of \cite
{Kac}.

%\subsection{\label{Remarks-Physics}On Physical Meaning}

%\texttt{TO BE ADDED : Observations on physical meanings, and the still
%elusive physical meaning of }$I_{6}$\texttt{...}

\subsection{\label{GSKG}On Special Geometry and \textit{``Generalized''}
Groups of Type $E_{7}$}

The sequence (\ref{N=2-mc}) and
\begin{equation}
\frac{SL\left( 2,\mathbb{R}\right) }{U\left( 1\right) }\times \frac{SO\left(
2,n-2\right) }{SO\left( 2\right) \times SO\left( n-2\right) },~n\geqslant 3,
\end{equation}
related to the case $m=2~$of (\ref{G4-red}), are the unique sequences of
symmetric non-compact spaces in the \textit{special K\"{a}hler geometry}
(SKG) of $\mathcal{N}=2$, $D=4$ vector multiplets (see \textit{e.g.} \cite
{magnific-7,CVP,SKG-Strominger,N=2-Big}, and Refs. therein).

Here we would like to discuss the characterization of SKG in terms of a
suitable \textit{``generalization'' }of the \textit{groups of type }$E_{7}$
\cite{Brown-Groups-of-type-E7} (for some preliminary discussion, see Sec. 4
of \cite{FMY-FD-1}).

As obtained in \cite{CFMZ1} (see Eq. (5.36) therein), the following real
function, which we dub \textit{``entropy functional''}, can be defined on
the vector multiplets' scalar manifold\footnote{%
Note that the expression (\ref{I4-N=2-symm}) is independent on the choice of
the symplectic frame and manifestly invariant under diffeomorphisms in $%
\mathbf{M}$.} $\mathbf{M}$:
\begin{equation}
\mathbb{I}_{4}=\left( \left| Z\right| ^{2}-Z_{i}\overline{Z}^{i}\right) ^{2}+%
\frac{2}{3}i\left( Z\overline{C}_{\overline{i}\overline{j}\overline{k}}Z^{%
\overline{i}}Z^{\overline{j}}Z^{\overline{k}}-\overline{Z}C_{ijk}\overline{Z}%
^{i}\overline{Z}^{j}\overline{Z}^{k}\right) -g^{i\overline{i}}C_{ijk}%
\overline{C}_{\overline{i}\overline{l}\overline{m}}\overline{Z}^{j}\overline{%
Z}^{k}Z^{\overline{l}}Z^{\overline{m}}.  \label{I4-N=2-symm}
\end{equation}
$Z$ is the central extension of $\mathcal{N}=2$, $D=4$ local supersymmetry
algebra, and $Z_{i}\equiv D_{i}Z$ are the so-called ``matter charges'' ($%
D_{i}$ stands for the K\"{a}hler-covariant differential operator; see
\textit{e.g.} \cite{CDF-rev} and \cite{N=2-Big} for notation and further
elucidation):
\begin{equation}
Z\equiv \mathcal{Q}^{M}V^{N}\mathbb{C}_{MN};~Z_{i}\equiv \mathcal{Q}%
^{M}V_{i}^{N}\mathbb{C}_{MN},  \label{Z-Zi-def}
\end{equation}
with $V^{M}$ denoting the vector of covariantly-holomorphic symplectic
sections of SKG, and $V_{i}^{M}\equiv D_{i}V^{M}$. Furthermore, $C_{ijk}$ is
the rank-$3$, completely symmetric, covariantly holomorphic tensor of SKG
(with K\"{a}hler weights $\left( 2,-2\right) $) (see \textit{e.g.} \cite
{CDF-1,CDF-2}):
\begin{equation}
\begin{array}{l}
C_{ijk}\equiv \mathbb{C}_{MN}\left( D_{i}D_{j}V^{M}\right) D_{k}V^{N}=-ig_{i%
\overline{l}}\overline{f}_{\Lambda }^{\overline{l}}D_{j}D_{k}L^{\Lambda
}=D_{i}D_{j}D_{k}\mathcal{S}=e^{K}W_{ijk}; \\
\overline{f}_{\Lambda }^{\overline{l}}\left( \overline{D}\overline{L}_{%
\overline{s}}^{\Lambda }\right) \equiv \delta _{\overline{s}}^{\overline{l}%
},~\mathcal{S}\equiv -iL^{\Lambda }L^{\Sigma }\text{Im}\left( F_{\Lambda
\Sigma }\right) ,~\overline{\partial }_{\overline{l}}W_{ijk}=0; \\
\overline{D}_{\overline{i}}C_{jkl}=0; \\
D_{[i}C_{j]kl}=0,
\end{array}
\label{C}
\end{equation}
the last property being a consequence, through the covariant holomorphicity
of $C_{ijk}$ and the SKG constraint on the Riemann tensor (see \textit{e.g.}
\cite{CDF-1,CDF-2,VP-what-SKG})
\begin{equation}
R_{j\overline{k}l\overline{m}}=-g_{j\overline{k}}g_{l\overline{m}}-g_{j%
\overline{m}}g_{l\overline{k}}+g^{i\overline{i}}C_{ijl}\overline{C}_{%
\overline{i}\overline{k}\overline{m}},  \label{SKG-constraint}
\end{equation}
of the Bianchi identities satisfied by the Riemann tensor $R_{i\overline{j}k%
\overline{l}}$.

Furthermore, $\mathbb{I}_{4}$ is an order-$4$ homogeneous polynomial in the
fluxes $\mathcal{Q}$; this allows for the definition of the $\mathcal{Q}$%
-independent rank-$4$ completely symmetric tensor $\Omega _{MNPQ}$ \cite
{FMY-FD-1}, whose general expression we explicitly compute here:
\begin{eqnarray}
\Omega _{MNPQ} &\equiv &2\frac{\partial ^{4}\mathbb{I}_{4}}{\partial
\mathcal{Q}^{(M}\partial \mathcal{Q}^{N}\partial \mathcal{Q}^{P}\partial
\mathcal{Q}^{Q)}}  \notag \\
&=&2V_{(M}V_{N}\overline{V}_{P}\overline{V}_{Q)}+2V_{i\mid (M}\overline{V}%
_{N}^{i}V_{j\mid P}\overline{V}_{Q)}^{j}-4V_{(M}\overline{V}_{N}V_{i\mid P}%
\overline{V}_{Q)}^{i}  \notag \\
&&+\frac{4}{3}i\left( \overline{C}_{\overline{i}\overline{j}\overline{k}%
}V_{(M}V_{N}^{\overline{i}}V_{P}^{\overline{j}}V_{Q}^{\overline{k}}-C_{ijk}%
\overline{V}_{(M}\overline{V}_{N}^{i}\overline{V}_{P}^{j}\overline{V}%
_{Q}^{k}\right)  \notag \\
&&-2g^{i\overline{i}}C_{ijk}\overline{C}_{\overline{i}\overline{l}\overline{m%
}}\overline{V}_{(M}^{j}\overline{V}_{N}^{k}V_{P}^{\overline{l}}V_{Q)}^{%
\overline{m}}.  \label{Omega-tensor} \\
&=&2V_{(M}V_{N}\overline{V}_{P}\overline{V}_{Q)}+2V_{i\mid (M}\overline{V}%
_{N}^{i}V_{j\mid P}\overline{V}_{Q)}^{j}-4V_{(M}\overline{V}_{N}V_{i\mid P}%
\overline{V}_{Q)}^{i}  \notag \\
&&+\frac{2}{3}\left( V_{(M}V_{N}^{\overline{i}}V_{P}^{\overline{j}}\overline{%
D}_{\overline{i}}\overline{V}_{\overline{j}\mid Q}+\overline{V}_{(M}%
\overline{V}_{N}^{i}\overline{V}_{P}^{j}D_{i}V_{j\mid Q}\right)  \notag \\
&&-2g^{i\overline{i}}\overline{V}_{(M}^{j}V_{N}^{\overline{l}}D_{i}V_{j\mid
N}\overline{D}_{\overline{i}}\overline{V}_{\overline{l}\mid Q)},
\label{Omega-tensor-2}
\end{eqnarray}
where the SKG defining relation (see \textit{e.g.} \cite
{CDF-1,CDF-2,VP-what-SKG})
\begin{equation}
D_{i}D_{j}V^{M}\equiv D_{i}V_{j}^{M}=iC_{ijk}\overline{V}^{k\mid M}
\label{SKG-rel-1}
\end{equation}
has been used in order to recast (\ref{Omega-tensor}) in terms of $V^{M}$, $%
V_{i}^{M}$ and $D_{i}V_{j}^{M}$ only.

Some further elaborations are possible; \textit{e.g.}, by using (\ref
{SKG-constraint}), $\mathbb{I}_{4}$ (\ref{I4-N=2-symm}) and $\Omega _{MNPQ}$
(\ref{Omega-tensor-2}) can respectively be rewritten as
\begin{eqnarray}
\mathbb{I}_{4} &=&\left| Z\right| ^{4}-\left( Z_{i}\overline{Z}^{i}\right)
^{2}-2\left| Z\right| ^{2}Z_{i}\overline{Z}^{i}+\frac{2}{3}i\left( Z%
\overline{C}_{\overline{i}\overline{j}\overline{k}}Z^{\overline{i}}Z^{%
\overline{j}}Z^{\overline{k}}-\overline{Z}C_{ijk}\overline{Z}^{i}\overline{Z}%
^{j}\overline{Z}^{k}\right) -\mathcal{R};  \label{I4-N=2-symm-2} \\
&&  \notag \\
\Omega _{MNPQ} &=&2V_{(M}V_{N}\overline{V}_{P}\overline{V}_{Q)}-2V_{i\mid (M}%
\overline{V}_{N}^{i}V_{j\mid P}\overline{V}_{Q)}^{j}-4V_{(M}\overline{V}%
_{N}V_{i\mid P}\overline{V}_{Q)}^{i}  \notag \\
&&+\frac{2}{3}\left( V_{(M}V_{N}^{\overline{i}}V_{P}^{\overline{j}}\overline{%
D}_{\overline{i}}\overline{V}_{\overline{j}\mid Q}+\overline{V}_{(M}%
\overline{V}_{N}^{i}\overline{V}_{P}^{j}D_{i}V_{j\mid Q}\right) -R_{MNPQ},
\label{Omega-tensor-3}
\end{eqnarray}
where the \textit{sectional curvature of matter charges} (\textit{cfr.} Eq.
(5.3) of \cite{Raju-1}; also note that (\ref{R-call-def}) is different from
the definition given by Eq. (3.1.1.2.11) of \cite{Kallosh-rev}))
\begin{equation}
\mathcal{R}\equiv R_{i\overline{j}k\overline{l}}\overline{Z}^{i}Z^{\overline{%
j}}\overline{Z}^{k}Z^{\overline{l}},  \label{R-call-def}
\end{equation}
and the corresponding rank-$4$ completely symmetric tensor
\begin{equation}
R_{MNPQ}\equiv \frac{\partial ^{4}\mathcal{R}}{\partial \mathcal{Q}%
^{(M}\partial \mathcal{Q}^{N}\partial \mathcal{Q}^{P}\partial \mathcal{Q}%
^{Q)}}=R_{i\overline{j}k\overline{l}}\overline{V}_{(M}^{i}V_{N}^{\overline{j}%
}\overline{V}_{P}^{k}V_{Q)}^{\overline{l}},  \label{R-sympl-tensor}
\end{equation}
have been introduced. Note that $R_{MNPQ}$ can be regarded as the completely
symmetric part of the \textit{``symplectic pull-back''} (through the
symplectic sections $V_{i}^{M}$) of the Riemann tensor $R_{i\overline{j}k%
\overline{l}}$ of $\mathbf{M}$.

Thus, SKG can be associated to a \textit{generalization} of the class of
groups of type $E_{7}$ \cite{Brown-Groups-of-type-E7}, based on $\mathbb{I}%
_{4}$ and the corresponding (generally field-dependent, non-constant) $%
\Omega $-structure:
\begin{equation}
\text{SKG~}:\left\{
\begin{array}{l}
\Omega _{MNPQ}:D_{i}\Omega _{MNPQ}=\partial _{i}\Omega _{MNPQ}\neq 0; \\
\\
\mathbb{I}_{4}\equiv \frac{1}{2}\Omega _{MNPQ}\mathcal{Q}^{M}\mathcal{Q}^{N}%
\mathcal{Q}^{P}\mathcal{Q}^{Q}\Rightarrow D_{i}\mathbb{I}_{4}=\partial _{i}%
\mathbb{I}_{4}\neq 0.
\end{array}
\right.  \label{gen-SKG}
\end{equation}

S\textit{ymmetric} K\"{a}hler spaces have a covariantly constant Riemann
tensor:
\begin{equation}
D_{i}R_{j\overline{k}l\overline{m}}=0.
\end{equation}
Within SKG, through the constraint (\ref{SKG-constraint}), this implies the
covariant constancy of the $C$-tensor (\ref{C}):
\begin{equation}
D_{(i}C_{j)kl}=D_{(i}C_{jkl)}=0,
\end{equation}
which in turn yields the relation:
\begin{equation}
C_{p(kl}C_{ij)n}g^{n\overline{n}}g^{p\overline{p}}\overline{C}_{\overline{n}%
\overline{p}\overline{m}}=\frac{4}{3}g_{\left( l\right| \overline{m}%
}C_{\left| ijk\right) }\Leftrightarrow g^{n\overline{n}}R_{\left( i\right|
\overline{m}\left| j\right| \overline{n}}C_{n\left| kl\right) }=-\frac{2}{3}%
g_{\left( i\right| \overline{m}}C_{\left| jkl\right) }.  \label{symm}
\end{equation}
Equivalently, \textit{symmetric} SK manifolds can be characterized by
stating that their $\Omega _{MNPQ}$ is is \textit{independent} on the scalar
fields themselves, and it matches the $\mathbb{K}$-tensor $\mathbb{K}_{MNPQ}$
defining the rank-$4$ invariant $\mathbb{K}$-structure of the corresponding $%
U$-duality group of type $E_{7}$ \cite{Brown-Groups-of-type-E7} (see also
\textit{e.g.} \cite{Exc-Reds}, and Refs. therein). Consequently, the
corresponding \textit{``entropy functional''} $\mathbb{I}_{4}$ (\ref
{I4-N=2-symm}) is \textit{independent} on the scalar fields themselves, and
it is thus a \textit{constant} function in $\mathbf{M}$, given by the $1$%
-centered limit of the \textit{Dixmier tensor} $I_{abcd}$ (\ref
{Dixmier-tensor-def}), which is nothing but the unique
algebraically-independent $1$-centered $U$-duality invariant polynomial $%
I_{4}$:
\begin{equation}
\underset{(U\text{-duality~group~}G_{4}\text{ is~of~type~}E_{7})}{\text{%
\textit{symmetric}~SKG}}\Rightarrow \left\{
\begin{array}{l}
\Omega _{MNPQ}=\mathbb{K}_{MNPQ}\Rightarrow D_{i}\Omega _{MNPQ}=\partial
_{i}\Omega _{MNPQ}=0; \\
\\
\mathbb{I}_{4}=I_{4}\equiv \frac{1}{2}\mathbb{K}_{MNPQ}\mathcal{Q}^{M}%
\mathcal{Q}^{N}\mathcal{Q}^{P}\mathcal{Q}^{Q}\Rightarrow D_{i}\mathbb{I}%
_{4}=\partial _{i}\mathbb{I}_{4}=0.
\end{array}
\right.  \label{symm-SKG}
\end{equation}
In turn, within \textit{symmetric} SKG, the pseudo-unitary $U$-duality group
$U\left( 1,s\right) $ (corresponding to $\mathcal{N}=2$ \textit{minimally
coupled} Maxwell-Einstein theory \cite{Luciani,Gnecchi-1}) is
``degenerate'', in the sense that the corresponding $I_{4}$ actually is the
square of the order-$2$ $U\left( 1,s\right) $-invariant polynomial $I_{2}$ (%
\ref{Inv-quadr}). Indeed, $\mathcal{N}=2$ \textit{minimally coupled}
supergravity is characterized by\footnote{%
By plugging $C_{ijk}=0$ into (\ref{SKG-rel-1}) and recalling definition (\ref
{Z-Zi-def}), one obtains the K\"{a}hler-covariant anti-holomorphicity of $%
V_{j}^{M}$, and thus of $Z_{j}$:
\begin{equation*}
D_{i}V_{j}^{M}=0\Rightarrow D_{i}Z_{j}=0.
\end{equation*}
} $C_{ijk}=0$, which plugged into (\ref{I4-N=2-symm}) (by taking (\ref
{symm-SKG}) into account) yields:
\begin{equation}
\underset{G_{4}=U\left( 1,s\right) }{\text{\textit{symmetric}~SKG~}}%
\Rightarrow \left\{
\begin{array}{l}
\Omega _{MNPQ}=\mathbb{K}_{MNPQ}\Rightarrow D_{i}\Omega _{MNPQ}=\partial
_{i}\Omega _{MNPQ}=0; \\
C_{ijk}=0; \\
\mathbb{I}_{4}=I_{4}=\left( \left| Z\right| ^{2}-Z_{i}\overline{Z}%
^{i}\right) ^{2}=\frac{1}{4}I_{2}^{2}\Rightarrow D_{i}\mathbb{I}%
_{4}=\partial _{i}\mathbb{I}_{4}=0,
\end{array}
\right.
\end{equation}
where the normalization of \cite{MS-FMO-1} (see Eq. (2.15) therein) has been
adopted.

We conclude by recalling that, as noticed in \cite{CFMZ1} and in \cite
{FMY-FD-1}, the \textit{``entropic functional'' }$\mathbb{I}_{4}$ (\ref
{I4-N=2-symm}) is related to the \textit{geodesic potential} defined in the $%
D=4\rightarrow 3$ dimensional reduction of the considered $\mathcal{N}=2$
theory. Under such a reduction, the $D=4$ vector multiplets' SK manifold $%
\mathbf{M}$ (dim$_{\mathbb{C}}=n_{V}$) enlarges to a \textit{special}
quaternionic K\"{a}hler manifold $\frak{M}$ (dim$_{\mathbb{H}}=n_{V}+1$)
given by $c$-map \cite{CFG,Ferrara-Sabharwal} of $\mathbf{M}$ itself : $%
\frak{M}=c\left( \mathbf{M}\right) $. By specifying Eq. (\ref{I4-N=2-symm})
in the ``$4D/5D$ \textit{special coordinates' ''} symplectic frame, $\mathbb{%
I}_{4}$ matches the opposite of the function $h$ defined by Eq. (4.42) of
\cite{dWVVP}, within the analysis of \textit{special} quaternionic
K\"{a}hler geometry. This relation can be strengthened by observing that the
tensor $\Omega _{MNPQ}$ given by (\ref{Omega-tensor})-(\ref{Omega-tensor-2})
is proportional to the $\Omega $-tensor of quaternionic geometry, related to
the quaternionic Riemann tensor by Eq. (15) of \cite{Bagger-Witten}; for
further comments, see \cite{FMY-FD-1}.

\section*{Acknowledgments}

A. M. gratefully acknowledges enlightening discussions with Laurent Manivel
and Bert Van Geemen, who also drew to his attention \cite{Kac} and \cite
{Dixmier}, respectively.

A. Y. would like to thank CERN Theory Division for kind hospitality.

The work of S. F. is supported by the ERC Advanced Grant no. 226455, \textit{%
``Supersymmetry, Quantum Gravity and Gauge Fields''} (\textit{SUPERFIELDS}),
and in part by DOE Grant DE-FG03-91ER40662.

The work of A. Y. is supported by the ERC Advanced Grant no. 226455, \textit{%
``Supersymmetry, Quantum Gravity and Gauge Fields''} (\textit{SUPERFIELDS}).

\end{document}